\documentclass[useAMS,usenatbib]{mnras}
\usepackage[pdftex]{graphicx} 
\usepackage{hyperref}
\usepackage{xcolor} 
\usepackage{lscape}
\usepackage{calc}
\usepackage{amsmath}
\usepackage{afterpage}
\usepackage{bm}
\usepackage{amssymb}

\hypersetup{colorlinks,linkcolor={red!50!black}, citecolor={blue!50!black},urlcolor={blue!80!black} }

\title[Star-forming protoclusters in simulations]
{Is there enough star formation in simulated protoclusters?}

\author[Lim et al.]
{Seunghwan Lim$^{1}$\thanks{E-mail: slim@phas.ubc.ca}\thanks{CITA National Fellow},
Douglas Scott$^{1}$,  
Arif Babul$^{2}$, 
David J. Barnes$^{3}$, 
Scott T. Kay$^{4}$, 
\newauthor
Ian G. McCarthy$^{5}$, 
Douglas Rennehan$^{2}$, 
Mark Vogelsberger$^{3}$ 
\\
\vspace*{6pt} \\
$^{1}$Department of Physics and Astronomy, University of British Columbia, Vancouver, BC, Canada V6T 1Z1 \\
$^{2}$Department of Physics \& Astronomy, University of Victoria, BC, V8X 4M6, Canada \\ 
$^{3}$Department of Physics, Kavli Institute for Astrophysics and Space Research, Massachusetts Institute of Technology, Cambridge, MA 02139, USA \\ 
$^{4}$Jodrell Bank Centre for Astrophysics, School of Physics and Astronomy, The University of Manchester, Manchester M13 9PL, UK  \\ 
$^{5}$Astrophysics Research Institute, Liverpool John Moores University, 146 Brownlow Hill, Liverpool L3 5RF, UK 
}

\begin{document} 

\pagerange{\pageref{firstpage}--\pageref{lastpage}}

\date{\today}
\pubyear{2020}

\maketitle

\label{firstpage}

\begin{minipage}{16.5cm}
\begin{abstract}\centering
\vspace{-4.5mm}
\begin{changemargin}{4.8cm}{0cm} 
\begin{minipage}{13cm}
As progenitors of the most massive objects, protoclusters are key to tracing the evolution and star-formation history of the Universe, and are responsible for ${\gtrsim}\,20$ per cent of the cosmic star formation at $z\,{>}\,2$. Using a combination of state-of-the-art hydrodynamical simulations and empirical models, we show that current galaxy-formation models do not produce enough star formation in protoclusters to match observations. We find that the star-formation rates (SFRs) predicted from the models are an order of magnitude lower than what is seen in observations, despite the relatively good agreement found for their mass-accretion histories, specifically that they lie on an evolutionary path to become Coma-like clusters at $z\,{\simeq}\, 0$. Using a well-studied protocluster core at $z\,{=}\,4.3$ as a test case, we find that star-formation efficiency of protocluster galaxies is higher than predicted by the models. We show that a large part of the discrepancy can be attributed to a dependence of SFR on the numerical resolution of the simulations, with a roughly factor of 3 drop in SFR when the spatial resolution decreases by a factor of 4. We also present predictions up to $z\,{\simeq}\,7$. Compared to lower redshifts, we find that centrals (the most massive member galaxies) are more distinct from the other galaxies, while protocluster galaxies are less distinct from field galaxies. All these results suggest that, as a rare and extreme population at high-$z$, protoclusters can help constrain galaxy formation models tuned to match the average population at $z\,{\simeq}\,0$. \\  \\
\textbf{Key\,\,words:} methods: statistical -- galaxies: formation -- galaxies: evolution -- galaxies: haloes -- submillimetre: galaxies -- galaxies: clusters: general \vspace{1cm}
\end{minipage}
\end{changemargin}
\end{abstract} 
\end{minipage}

%\end{abstract} 

%\begin{keywords} 
%methods: statistical -- galaxies: formation -- galaxies: evolution -- galaxies: haloes -- submillimetre: galaxies -- galaxies: clusters: general 
%\end{keywords}

\vspace{-1.4cm}

%%%%%%% SECTION 1
\section[intro]{INTRODUCTION}
\label{sec_intro}

Protoclusters of galaxies, defined as the high-redshift progenitors of galaxy clusters, are key to understanding the evolution of the early Universe. Since galaxy clusters are the most massive virialized objects, they are expected to assemble and emerge latest in the picture of hierarchical structure growth \citep[e.g.,][]{sheth1999,mo2010}. However, observations have accumulated a vast amount of evidence for `archaeological' downsizing, whereby more than 20 per cent (and up to 50 per cent) of the cosmic \vspace{5mm}
\\ \\ 
* E-mail: slim@phas.ubc.ca \\
$\dagger$ CITA National Fellow \vspace{1cm} \\ 
star formation occurred in galaxy clusters at $z\,{\simeq}\,2$, in contrast to a fairly negligible contribution ($<1$ per cent) at $z\,{\simeq}\,0$ \citep[e.g.,][]{dressler1980,kauffmann2004,stanford2006,vanderburg2013,madau2014,chiang2017}. This underlines the importance of protoclusters in understanding the formation and evolution of galaxies (and particularly massive ones) and large-scale structure \citep[e.g.,][]{overzier2016, oteo2018, rennehan2020}. 

A great amount of effort has been put into identifying and studying high-redshift protoclusters. Whereas clusters at lower redshifts are usually detected via their Bremsstrahlung radiation emitted in the X-ray from their cores \citep[e.g.][]{rosati2009,wang2016,mantz2018} or via the Sunyaev-Zeldovich effect \citep[e.g,][]{planckxx2014,huang2020}, the same detection methods are less useful for identifying {\it proto\/}clusters because the signals typically fall below current instrument sensitivity. Instead, if one is interested in observing protoclusters during their peak of star formation, one can look for overdense regions of galaxies that are mostly star forming, which we see at (sub-)millimetre wavelengths from their dust-reprocessed stellar emission \citep[e.g.][]{steidel2000,venemans2007,miley2008,casey2014,chiang2013, chiang2014,miller2015, toshikawa2016, lovell2018}. A number of protoclusters have been detected in this way, and their total star-formation rates (SFRs) typically reach up to around several thousands of ${\rm M_\odot\,yr^{-1}}$ \citep[e.g.][]{ivison2013, dannerbauer2014, umehata2014, umehata2015, yuan2014, flores-cacho2015, hung2016, wang2016, miller2018, hill2020, Rotermund}. 

Comparison between observed protoclusters and theoretical predictions have been attempted before, but the findings in most (if not all) cases is that the SFRs in simulated protoclusters at high-$z$ fall significantly short (typically by a factor of 2--5) of matching the observations, while simulations predict much higher SFRs and stellar masses for clusters at low redshifts than obserations \citep[e.g.][]{granato15, bassini20}. \citet{lovell20} found about a factor of 2 deficit in their simulations relative to observations of sub-mm source number counts in single-dish surveys (although their simulations only had a side-length of ${\lesssim}\,150\,{\rm Mpc}$, which is too small to contain even a single protocluster with SFRs as extreme as reported in the literature, as we discuss later). However, the origin of the discrepancy and its evolution with redshift remains unclear. Another issue that is at least partly related is the long-standing challenge for models and simulations to reproduce the observed properties of main-sequence star-forming galaxies at high redshifts \citep[e.g.][]{granato00, baugh05, fontanot07, shimizu12, dave2019, donnari19, mcalpine19, hayward20}, although \citet{hayward2013a} and \citet{cowley15} improved the match to observations significantly. The reason that this is believed to be connected to the same issue for (proto)clusters is because observations have found no noticeable difference in SFRs between individual sub-mm galaxies (SMGs) in (proto)cluster and those in other environments \citep[e.g.][]{Rotermund}. While such a conclusion may need to be further investigated with a larger sample size, if it continues to hold true, then the two challenges would appear to be connected. 

Comparing these star-forming protoclusters with simulations presents several critical challenges. First, the typical box size of current cosmological hydrodynamical simulations is not large enough to contain a sufficient number of the most massive objects in the Universe for a statistically reliable analysis of the average population, nor to contain possible outliers. The number density of (proto)clusters at $z\,{>}\,2$ is expected to be about one per every $10^7$ to $10^9\,{\rm Mpc}^3$ (even this range is highly uncertain due to completeness)\footnote{For instance, \citet{casey2016} estimated the number density of DSFG-rich protoclusters to be $\simeq 5\times10^{-8}\,{\rm Mpc^{-3}}$ while \citet{wang2020} estimated it to be $10^{-10}\,{\rm Mpc^{-3}}$ from their sample of nine SPT protocluster candidates. For reference, the number density of massive clusters at $z\,{=}\,0$ with $M_{\rm cl}\,{\simeq}\,10^{15}\,{\rm M_\odot}$, the probable descendants of protoclusters when assuming an average mass evolution, is estimated to be $\lesssim 10^{-8}\,{\rm Mpc^{-3}}$ \citep[e.g.][]{bahcall93}.}, while most hydrodynamical cosmological simulations are run on a periodic box whose size is, at best, comparable to this volume, thus containing only a few such systems \citep{dubois2014, khandai2015, schaye2015, bocquet2016, dave2016, mccarthy2017, pillepich2018, dave2019}. The fact that observations are also not perfectly free from this limitation further complicates the comparison. Secondly, SFR enhancements from galaxy interactions may be underestimated in large-volume simulations due to insufficient numerical resolution \citep[e.g.][]{sparre2015, sparre2016, grand2017}. Thirdly, it is generally not straightforward to make fair comparisons between simulated data and direct observables because of observational effects, such as selection criteria and the uncertain conversion of observables to physical properties. Finally, both predictions from simulations and the interpretation of observational data at high redshift rely strongly on assumptions that have only been confirmed to be valid at lower redshift, such as the use of an initial mass function, dust attenuation and conversion factors inferred from low-$z$ observations. Therefore it is unclear whether discrepancies reported from recent studies between simulations and observations, such as the discrepancy in the number density of protoclusters \citep[e.g.][]{miller2015,casey2016}, arise due to the uncertainties listed above or are due to more fundamental failures in the models used to make the predictions. Nonetheless, models generally agree on the downsizing of star formation in massive objects \citep[e.g.][]{hayward2013a, rennehan2020}.

In this paper, we use a high-resolution hydrodynamical simulation, IllustrisTNG, and the merger tree derived from it, together with a suite of zoom simulations of $390$ massive clusters, MACSIS, to study the discrepancies between simulated and observed protoclusters in detail. Section~\ref{sec_model} describes the simulations and models that we use for our analysis, while Sect.~\ref{sec_obs} summarizes the observational data. In Sect.~\ref{sec_methods}, we describe how we specifically make comparisons between the models and data, and Sect.~\ref{sec_results} presents our results. Section~\ref{sec_spt2349} examines in further detail a particular case study of a protocluster, SPT2349$-$56, for which recent observations inferred an exceptionally high SFR at $z\,{=}\, 4.3$ \citep{miller2018, hill2020}. We discuss our findings in Sect.~\ref{sec_discussion}, and summarize and conclude in Sect.~\ref{sec_sum}.

%%%%%%% SECTION 2
\section[models]{MODELS}
\label{sec_model}

Here we describe the models that we choose to compare with observation. Recently, there have been improvements to hydrodynamical simulations, yielding a better match to observational data, mainly due to better prescriptions for subgrid models of stellar and active galactic nucleus (AGN) feedback \citep{schaye2015,mccarthy2017,weinberger2017,pillepich2018} and for the mixing of the resulting energy, momentum and metals \citep[e.g.][]{rennehan2019}. However, hydrodynamical simulations still face a number of challenges. 

A major issue is numerical convergence: most large-volume simulations with the highest resolution attained so far still fail to reach numerical convergence for many specific observables \citep[see e.g.][]{sparre2015, sparre2016, grand2017,pillepich2018,vogelsberger2018}. The problem mainly arises from the fact that the baryonic physics governing star formation and feedback operate on the roughly parsec scale, while the best resolution currently achieved in simulations reaches only down to around $50\,{\rm pc}$ \citep[e.g.][]{guedes2011, wang2015, sparre2015, sparre2016, sawala2016, tremmel2017, tremmel2019}. The Feedback In Realistic Environments (FIRE) project \citep{hopkins2014, wetzel2016} probes down to a smaller scale of 1--10\,pc, but their simulation suites, as well as many of the simulations listed above, are limited to haloes less massive than $10^{14.5}\,{\rm M_\odot}$ \citep[MassiveFIRE;][]{feldmann2016, feldmann2017}, and are thus yet to be suitable for studying protoclusters \citep[see table~2 of][for a review of current simulations]{vogelsberger2020}. In addition, the parameters of those subgrid models are typically tuned to match low-$z$ observations, making it hard to break degeneracies among models of different evolutionary paths that end up with similar $z\,{\simeq}\,0$ properties (studies of high-redshift comparisons like this study are thus critical to constrain the detailed choices within such models). Additionally, it is not always obvious how to make like-with-like comparisons between data extracted from simulations and direct observables, due to observational selection effects. Finally, numerical simulations are computationally expensive to run, prohibiting the exploration of a large parameter space of physics recipes.

In order to overcome those caveats of simulations, there have been alternative approaches to associate galaxies and haloes using a simple modelling procedure, usually via a parameterization that describes the relation between dark matter haloes and their stellar components across redshifts \citep[e.g.][]{jing1998,vandenbosch2003,bower2006,croton2006,behroozi2013,moster2013,lu2014,lim2017}. However, this approach still often relies on simulations indirectly, via merger trees that are extracted from simulations and used to trace the mass-accretion history (MAH) of dark matter haloes. Dark-matter-only simulations and the derived MAHs are much more robust against the effects of numerical resolution than full hydrodynamical simulations, and are also at this point quite well understood. Additionally, such approaches are computationally inexpensive, and can use direct observables from a wide range of redshifts as constraints for the models. Nevertheless, because each astronomical survey observes with different instrumentation and observing techniques, and observes different populations of galaxies at various evolutionary stages across time, it is difficult to make a compilation of observational data that is consistent in terms of systematics and limitations -- this makes it hard to evaluate uncertainties in the models constrained by the data. Moreover, it is almost impossible to make accurate predictions for individual galaxies from these types of procedures, since the models are tuned based on average populations at given mass and redshift. More importantly, model predictions highly depend on their parameterizations and how exactly they are tuned. 

Therefore, to capture the whole range of limitations in models, we choose to include predictions from both simulations and from some empirical models for our comparison with the observational data. In the next subsections we provide short descriptions of each model, and refer the reader to the original papers and references therein for more details. 

%%%% Table 1 %%%%
\begin{table*}
 \renewcommand{\arraystretch}{1.4} 
 \centering
 \begin{minipage}{102mm}
  \caption{List of simulations used. From left to right the columns show: simulation name; comoving box size; baryonic mass; dark matter particle mass; and the Plummer-equivalent gravitational softening length.}
  \begin{tabular}{ccccc}
\hline
Simulation & $L$ & $m_{\rm b}$ & $m_{\rm DM}$ &  $\epsilon_{\rm DM}$ \\
 & $[{\rm cMpc}]$ & $[h^{-1}\,{\rm M_\odot}]$ & $[h^{-1}\,{\rm M_\odot}]$ &  \\
\hline
\hline
TNG300-1 & 303 & $7.6\times10^6$ & $4.0\times10^7$  & $2\,h^{-1}{\rm ckpc}\rightarrow 1\,h^{-1}{\rm kpc}$\textsuperscript{\footnote{For TNG300, the softening length is fixed in comoving units for $z\,{>}\,1$ and in physical units thereafter.}} \\  
TNG300-3 & 303 & $4.8\times10^8$ & $2.5\times10^9$  & $8\,h^{-1}{\rm ckpc}\rightarrow 4\,h^{-1}{\rm kpc}$ \\  
MACSIS & 3200\textsuperscript{\footnote{The box size of the parent simulation.}} & $8.0\times10^8$ & $4.4\times10^9$  & $4\,h^{-1}{\rm ckpc}\rightarrow 4\,h^{-1}{\rm kpc}$\textsuperscript{\footnote{For MACSIS, the softening length is fixed in comoving units for $z\,{>}\,3$ and in physical units thereafter.}} \\  
\hline
\\
\vspace{-9mm}
\end{tabular}
\textbf{Notes.}
\vspace{-5mm}
\label{tab_sim}
\end{minipage}
\vspace{0mm}
\end{table*}
%%%%%%%%%%%%%%%%%%%%%%%%%%%%%%%%%%%%%%%%%%%%%%%%%%%%%%%%%%%%%%%%%%%%%%%%%%%%%%

\subsection{TNG300 simulations}
\label{ssec_TNG}

The IllustrisTNG project \citep[that we will refer to as `TNG';][]{marinacci18, naiman18, nelson18, nelson19, pillepich18b, springel18} has produced a series of cosmological hydrodynamical simulations. We choose TNG for our analysis because, as will be shown later, star formation in protoclusters strongly depends on resolution and TNG has one of the highest resolutions for its cosmological volume, as well as having a wide range of resolutions, which allows us to explicitly probe the effect of resolution. TNG uses the \textsc{Arepo} code \citep{springel10}, with significantly updated schemes applied to magnetohydrodynamics \citep{pakmor11} and `wind-mode' black-hole feedback, in particular, compared to the original Illustris project \citep{vogelsberger14a, vogelsberger14b, genel14, sijacki15}, the predecessor of TNG. The simulations assume the cosmological parameters of \citet{pcxiii}, with $\sigma_8\,{=}\,0.816$, $h\equiv H_0/(100\,{\rm km}\,{\rm s}^{-1}\,{\rm Mpc}^{-1})\,{=}\,0.677$, $\Omega_{\rm m}\,{=}\,0.309$ and $\,\Omega_{\rm b}\,{=}\,0.0486$. 

At the time of this paper being written, the TNG outputs have been made public for two cosmological volumes, which are approximately $100\,{\rm cMpc}$ and $300\,{\rm cMpc}$ on a side, where ${\rm cMpc}$ is a comoving Mpc unit (as opposed to a physical Mpc, which we call ${\rm Mpc}$). Because the objects we are interested in are extremely rare, we exclusively use the latter simulation for analysis in this paper, and refer to this as `TNG300'. TNG300 is provided at three different resolutions, referred to as `TNG300-1', `TNG300-2' and `TNG300-3', in order of high to low resolution. In TNG300-1, the periodic box of $(300\,{\rm cMpc})^3$ is sampled with $(2500)^3$ dark matter particles and $(2500)^3$ gas particles, with the target baryon and dark matter particle mass being $m_{\rm b}=1.1\times10^7\,{\rm M_\odot}$ and $m_{\rm DM}=5.9\times10^7\,{\rm M_\odot}$, respectively. The $z\,{=}\,0$ Plummer-equivalent gravitational softening length for the collisionless component ($\epsilon^{z=0}_{\rm DM,\ast}$) is $1.5\,{\rm kpc}$, and the minimum comoving value of the adaptive gas gravitational softening ($\epsilon_{\rm gas}^{\rm min}$) is $0.25\,h^{-1}{\rm ckpc}$. TNG300-2 (TNG300-3) is run with $8$ ($64$) times lower mass resolution for both the dark matter and gas components, with the softening lengths twice (four times) those of TNG300-1. The analysis and results presented in this paper are mainly based on TNG300-1, and hereafter we refer to TNG300-1 as `TNG300' for the sake of brevity, except where stated explicitly (e.g.\ in Sect.~\ref{sec_discussion}, where we investigate the impact of numerical resolution). 

The snapshots from TNG are saved and provided at a total of 100 epochs, from $z\,{\simeq}\, 20$ to $z\,{=}\,0$, with the timesteps between the snapshots ranging between 50 and 200\,Myr. Later in our analysis, we will be choosing the snapshot closest to the redshift relevant for the observational data that we compare to. 

Because the masses and spatially resolved length sizes are larger than the typical scales of the interstellar medium (ISM) where star formation takes place (even for TNG300-1), star formation is treated using a subgrid model. Specifically, the simulations follow \citet{springel2003} for the star-formation prescription and for pressure within the multi-phase ISM. In this recipe, gas with density above a threshold of $n_{\rm H}\,{\simeq}\,0.1\,{\rm cm}^{-3}$ forms stars at a rate that reproduces the Kennicutt-Schmidt law \citep{schmidt1959,kennicutt1998} in a stochastic manner, with a characteristic gas consumption timescale of $t_{\rm SFR}=2.2\,{\rm Gyr}$ at the threshold density. The simulations specifically assume a Chabrier initial mass function \citep[IMF;][]{chabrier2003}. 

From each snapshot, haloes are identified by running the friends-of-friends \citep[FoF;][]{davis1985} algorithm only on the dark matter particles, with a linking length of $0.2$ times the average separation between the particles. The other types of particles, such as gas and stars, are assigned to the same FoF halo as their nearest dark matter particle. For each halo, we define a total mass, $M_{200}$, by accounting for all types of halo particles within $R_{200}$.\footnote{$R_{200}$ is the radius within which the mean enclosed mass density is $200$ times the critical density of the Universe, and $M_{200}$ is the mass enclosed within $R_{200}$.}. Definitions of mass used for our analysis are detailed in Sect.~\ref{sec_methods}. After this step, the substructures of haloes (also referred to as `subhaloes'), where the galaxies reside, are identified using the \textsc{Subfind} \citep{springel2001,dolag2009} algorithm. 

\subsection{Merger tree of TNG300}
\label{ssec_tree}

Although we do not have sufficient understanding of the observed protoclusters to match populations across time and to compare their detailed evolution with the simulations, it may still be useful to investigate the merger trees of clusters from simulations in order to understand how they build up their stellar mass, and look for any discrepancies with observation within specific snapshots. 

To this end, we use the merger tree from TNG300. Two distinct merger trees have been generated and released for TNG300, using \textsc{SubLink} \citep{rodriguez-gomez2015} and \textsc{LHaloTree} \citep{springel2005}. We find that our analysis and conclusions are insensitive to which of the two merger trees are used. We choose to present our results based on the \textsc{SubLink} trees throughout the paper because these trees allow us to track objects in a more consistent way for our specific science goal, with provided links to `descendants', `first and last progenitors', `root descendants' and `main leaf progenitors', as well as direct links to (sub)haloes and their properties at each snapshot -- all useful for tracing the evolution of (proto)clusters. The details of how we specifically use the merger trees for our analysis are presented in Sect.~\ref{sec_methods}.

\subsection{MACSIS simulations}
\label{ssec_MACSIS}

Carrying out an effective simulation of protoclusters is challenging.  Given the large scales involved in their collapse process, covering a few tens of cMpc, a large volume needs to be simulated, while high resolution is also required to accurately capture the physics involved in their highly dense environments. At the same time, however, since they are quite rare, it is not guaranteed to have protoclusters form in a simulated box if the region and initial conditions for the box are chosen randomly. With these considerations, it may be more appropriate to study protoclusters using `zoom' simulations, where simulations are re-run at much higher resolution and in a smaller sub-volume that is specific to the formation of the objects of interest, derived from a parent simulation that is large enough to contain a reasonable number of such objects. Because only the selected objects from a parent simulation are re-run, however, it may bias results if a selection for a re-simulation does not match that from observation for a given study. 

We examine the results from the Virgo consortium's MAssive ClusterS and Intercluster Structures (MACSIS) project \citep{barnes2017} in our comparisons with the observational data. MACSIS is a suite of 390 massive clusters selected from a large `parent' dark matter-only simulation and resimulated with full baryonic physics (i.e.\ hydrodynamics and prescriptions for star formation and feedback). Because MACSIS adopts mostly the same baryonic physics as applied in the BAryons and HAloes of MAssive Systems (BAHAMAS) simulations \citep{mccarthy2017}, it can be considered as an extension of BAHAMAS to the most massive (and rare) collapsed objects. 

The parent model here is a dark matter-only simulation that is $3.2\,{\rm cGpc}$ on a side, using cosmological parameters explicitly from \citet{pcxiii}, the values of which are $\sigma_8\,{=}\,0.829,\, h\,{=}\,0.678,\, \Omega_{\rm m}\,{=}\,0.307$ and $\Omega_{\rm b}\,{=}\,0.0483$. We refer the reader to \citet{barnes2017} for a more detailed description of the parent simulation. 

From a total of $9754$ massive haloes (${>}\,10^{15}{\rm M_\odot}$ at $z\,{=}\,0$) that are identified in the parent simulation, a sample of $390$ massive haloes was randomly selected for the high-resolution resimulation. With logarithmically spaced bins of $\Delta \log_{10}M_{\rm FoF}=0.2$, all haloes were selected from bins containing less than $100$ haloes. For any bin containing more than $100$ haloes, the bin was subdivided into bins of $0.02\,$dex, from each of which $10$ haloes were randomly chosen for the resimulation. The resulting sample of $390$ haloes is summarized in Table~1 of \citet{barnes2017} and is found to represent well the underlying population of the parent simulation in their properties. 

This sample of haloes was resimulated using a zoom simulation technique \citep{katz1993,tormen1997}. The mass of particles in the zoom simulations are $m_{\rm DM}=4.4\times 10^9\,h^{-1}{\rm M_\odot}$ and $m_{\rm gas}=8.0\times 10^8\,h^{-1}{\rm M_\odot}$, and the Plummer equivalent softening length is from $4\,h^{-1}{\rm ckpc}$ (at $z\,{>}\,3$) to $4\,h^{-1}{\rm kpc}$. Note that the mass resolution and softening length match those of BAHAMAS, and are comparable to those of TNG300-3, while the mass resolution is much lower than that of TNG300-1. 

We refer the reader to \citet{schaye2010}, \citet{lebrun2014} and \citet{mccarthy2017} for details of the subgrid models adopted in MACSIS, and only provide a brief summary here. In MACSIS, star formation is implemented following the basic recipe of \citet{schaye2008}, such that gas particles with a density of $n_{\rm H}\,{>}\,0.1\,{\rm cm}^{-3}$ form stars stochastically at a pressure-dependent rate that matches the Kennicutt-Schmidt law, assuming a Chabrier IMF. Radiative cooling is modelled based on \citet{wiersma2009a}, while stellar evolution and chemical gas enrichment are computed following \citet{wiersma2009b}. For stellar feedback, the kinetic wind model of \citet{dallavecchia2008} is implemented. The simulation follows \citet{booth2009} for the seeding and growth of black holes, as well as for feedback driven by black holes. 

\subsection{Empirical models}
\label{ssec_model}

Another way to explore the star-formation history of haloes is via modelling of the stellar mass versus halo mass relation (SMHMR) or via modelling of the SFR in haloes of a given property using a simple parameterization, where the parameter values are constrained by observational data. Significant progress has been made in recent years with these `empirical' models, which link the evolution of galaxies to the evolution of dark matter haloes. This approach generally begins by assuming a parameterized SMHMR that is allowed to vary with redshift through another parameterized function. Then the evolution of stellar mass is traced by applying the SMHMR to merger trees of dark matter haloes extracted from either simulations or a merger tree generator \citep[e.g.][]{parkinson2008}. The difference in stellar mass between each timestep of the merger tree (after accounting for the changes due to mass loss and mergers) is taken as the in situ star formation between the timesteps, and the quotient of this and the step size is used to estimate the SFR. The resulting stellar component of galaxies computed for each set of parameter values is compared with observations to find the parameters that match the observations best. Here we adopt two specific models, namely those of \citet{behroozi2013} and \citet{moster2013}. 

Following the general approach described above, \citet[][hereafter B13]{behroozi2013} constrained the SMHMR using observational constraints of the stellar mass function (SMF), specific SFR (sSFR) and cosmic SFR density (CSFD) at various redshifts. In particular, they made use of a new compilation of data, including observations at very high redshifts \citep{bouwens2011,mclure2011,bradley2012}, which enabled them to place constraints on the model to $z\,{\simeq}\, 8$, while previous works only used observations typically at $z\,{<}\,2$, limiting the validity of models significantly. B13 also accounted for numerous observational systematics, such as the measurement error in stellar mass due to uncertainties in stellar population synthesis, dust models and star-formation histories (SFHs), as well as scatter in the SMHMR and incompleteness, by including them in the parameterization and constraining them together with the other parameters. Aside from the parameters set to account for the uncertainties, the SMHMR in their modelling consists of five components: a characteristic halo mass; a characteristic ratio; the faint-end slope; and the massive-end slope and amplitude. Each of these is allowed to vary with redshift through additional functional parameters. We note that \citet{behroozi2019} updated this B13 model with a new, more comprehensive set of observational data as constraints, but the change in the results is insignificant for our analysis. 

Another empirical model that we consider for comparison is that of \citet[][hereafter M13]{moster2013}. Unlike B13, where observed SFRs were also used as constraints, M13 used only a compilation of observed SMFs at $0\,{<}\,z\,{<}\,4.5$ to constrain their model, and compared predictions of sSFR and CSFD with those from observation to validate the model. Their parameterization of the SMHMR is a double power law, thus consisting of four parameters: the characteristic halo mass; the normalization; and the slopes for the low-mass and high-mass ends of the SMHMR. Each of these parameters are further broken down to track their evolution with redshift. Similar to B13, M13 incorporates in their model the measurement error of the observed stellar mass. The CSFDs and sSFRs for haloes with different masses predicted by the model thus constrained are found to match those from observations reasonably well, except for massive systems at redshifts $z\,{\gtrsim}\,2$, which is the regime that our study focuses on. The model of M13 was recently updated in \citet{moster2018} by directly relating the SFR (instead of stellar mass) to the MAH, and by using an updated compilation of observations including the sSFR and CSFD as additional constraints. We find, however, that within the uncertainties the SFRs predicted from the updated results agree well with those from M13.

%%%%%%% SECTION 3
\section[data]{OBSERVATIONAL DATA} 
\label{sec_obs}

Galaxy protoclusters at high redshifts are not observable through standard cluster-detection techniques, such as thermal X-rays and the Sunyaev-Zeldovich effect, since the signals usually fall below the current detection limits of instruments. Instead, the main observational signature of galaxy protoclusters comes from the galaxies themselves, which emit light primarily at infrared, (sub)-millimetre and radio wavelengths. 

Galaxy protoclusters can be discovered via their stellar emission by searching for overdensities of galaxies, either photometrically or spectroscopically, in large-area surveys. Examples of such protoclusters include the SSA22 structure \citep[][$z\,{\simeq}\,3.1$]{steidel2000,umehata2015}, the Subaru Deep Survey structure \citep[][$z\,{\simeq}\,4.9$]{shimasaku2003}, the COSMOS $z\,{=}\,2.44$ structure \citep{chiang2015}, PC 217.96$+$32.3 \citep[][$z\,{\simeq}\,3.8$]{dey2016}, and z57OD and z66OD in the SXDS field \citep[][$z\,{\simeq}\,6$]{harikane2019}.

Galaxies residing in protoclusters are also expected to be very active and have large star-formation rates \citep[e.g.][]{rennehan2020}. A large star-formation rate corresponds to a large amount of far-infrared emission (which we see at submillimetre-to-millimetre wavelengths in the observed frame), and many galaxy protoclusters have been discovered by searching for similar overdensities of submillimetre galaxies, such as the GOODS-N protocluster \citep[][$z\,{=}\,1.99$]{chapman2009}, the COSMOS $z\,{=}\,2.10$ cluster \citep{yuan2014,hung2016}, PCL1002 \citep[][$z\,{=}\,2.44$]{casey2015} and the Distant Red Core \citep[`DRC', at $z\,{\simeq}\,4.0$;][]{oteo2018,long2020}. Another approach is to search for bright, unresolved peaks in large-area millimetre or submillimetre surveys, and then to follow up promising targets with higher-resolution facilities \citep[e.g.][]{clements14}. Data from cosmic microwave background experiments, for example the all-sky {\it Planck\/} survey \cite{PlanckHerschel,PHz}, are particularly useful for this purpose.  This technique has uncovered protoclusters such as PHz G95.5$-$61.6 \citep[][$z\,{\simeq}\,2$]{flores-cacho2015}, PHz G073.4$-$57.5 \citep[][$z\,{=}\,1.5$--$2.4$]{kneissl2019}, and SPT2349$-$56 \citep[][$z\,{=}\,4.3$]{miller2018,hill2020}; indeed, the South Pole Telescope (SPT) survey has uncovered nine protocluster candidates to date at $z\,{=}\,3$--$7$ \citep{wang2020}.

Another feature of actively star-forming galaxies is that they might contain AGN, which can be detected at even longer wavelengths. Indeed, radio galaxies have been used as signposts for uncovering protoclusters, helping discover objects around the AGN HS1700$+$643 \citep[][$z\,{\simeq}\,3.1$]{steidel2000}, HS1549$+$19 \citep[][$z\,{\simeq}\,2.6$]{steidel2011} and MRC1138$-$262 \citep[][$z\,{=}\,2.16$]{dannerbauer2014}.

Given the wide range of selection techniques used to find and catalogue galaxy protoclusters, comparison between samples is challenging. On the other hand, differences between selection techniques could also be used to trace protoclusters at different stages on their evolutionary path towards clusters, and hence fill in the picture of how large-scale structure forms. In particular, the use of submillimetre and millimetre wavelengths provides a direct probe of star formation, giving us a technique that is biased towards the epoch of maximal star-formation activity, rather than tracing stellar mass, virialized mass or IGM gas.

Because we are interested in the star formation within protoclusters, we construct an observational sample of star-formation-selected protoclusters to be compared with the models in our analysis. Specifically, the sample that we use consists of the seven structures from \citet{casey2016}, i.e. the GOODS-N protocluster, the COSMOS $z\,{=}\,2.10$ cluster, MRC1138$-$262 and the COSMOS (at $z\,{=}\,2.44$), SSA22 (at $z\,{=}\,3.1$), HDF\,850.1 (at $z\,{=}\,5.18$) and AzTEC-3 at ($z\,{=}\,5.30$) structures.\footnote{We do not include the GN20 overdensity \citet{casey2016} because of the large uncertainty in its volume estimate, which is required for conversions to quantities within $5R_{500}$ later in our analysis.} The masses of these structures are based on the same paper and their SFRs come from \citet{hill2020}. We also compare with eight more SPT protocluster candidates with properties based on \citet{wang2020}, namely SPT0303$-$59 ($z\,{\simeq}\,3.3$), SPT0311$-$58, SPT0348$-$62 ($z\,{\simeq}\,5.7$), SPT0457$-$49 ($z\,{\simeq}\,4.0$), SPT0553$-$50 ($z\,{\simeq}\,5.3$), SPT2018$-$45 ($z\,{\simeq}\,3.2$), SPT2052$-$56 ($z\,{\simeq}\,4.3$) and SPT2335$-$53 ($z\,{\simeq}\,4.8$). Finally we include the DRC and SPT2349$-$56, with their properties based on \citet{oteo2018} and \citet{long2020} and on \citet{miller2018}, \citet{hill2020} and \citet{Rotermund}, respectively. Note that we make this sample selection because these are well-studied structures with properties that are relatively well constrained. Some of the properties explored in our analysis are available only for a subset of this sample from the literature, in which case we only use the appropriate subset for a comparison to our results.

%%%%%%% SECTION 4
\section[methods]{METHODS}
\label{sec_methods}

%%%%%%%%%%% figure
\begin{figure*}
\includegraphics[width=1.0\linewidth]{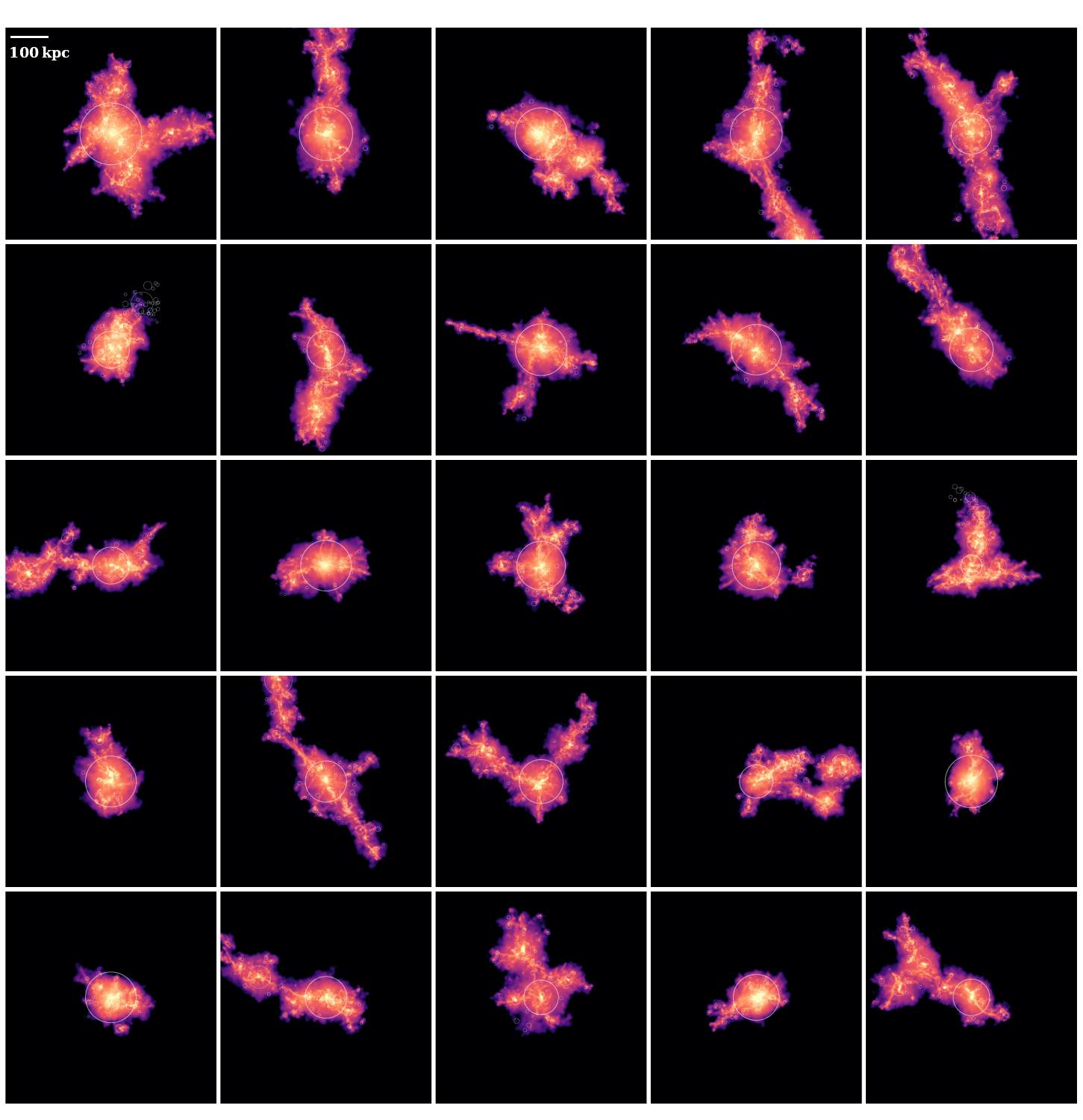}
\caption{Distribution of the gas cells associated with the main halo for the 25 most massive protoclusters (i.e. the progenitors of the 25 most massive $z=0$ clusters; see the text for a detailed definition of a protocluster and its main halo) from the TNG300 simulation at a redshift of $4.4$. The particles are colour-coded according to their star-formation rate (SFR) on a logarithmic scale. The largest circle in each panel shows the location of $R_{500}$ for the main halo, while the smaller circles indicate subhaloes. }
\label{fig_TNGsample}
\end{figure*}

The objects of particular interest to us here are the most massive in the Universe, which are expected to evolve to become at least Virgo-like clusters (more likely up to Coma cluster-type) at $z\,{=}\,0$, with total mass of $M_{200}\simeq 10^{15}\,{\rm M}_\odot$. TNG300 contains approximately 50 (25) simulated clusters with $M_{\rm 200}\,{>}\,10^{14.5}\,{\rm M}_\odot$ (${>}\,10^{14.6}\,{\rm M}_\odot$) at $z\,{=}\,0$. From TNG300, we choose the $25$ most massive clusters identified at $z\,{=}\,0$ for our analysis. This is nearly the maximum number of objects available from the simulation within the estimated mass range of observed protoclusters, and at the same time is a large enough number to study their average trends (although average trends from the mass-limited sample could be biased with respect to the mainly SFR-based selection from the observations, as will be discussed later), as well as to include potential outliers of this already extreme population. The total mass of the TNG300 sample ranges from $M_{200}= 10^{14.56}\,{\rm M}_\odot$ to $10^{15.19}\,{\rm M}_\odot$, with a median of $10^{14.70}\,{\rm M}_\odot$. 

From the MACSIS simulations, on the other hand, we use the entire sample of 390 clusters (as described in Sect.~\ref{ssec_MACSIS}) for our analysis. The MACSIS sample complements that of TNG300, with a much larger sample size, increasing the chance of having extreme outliers in particular, but also focuses on even more massive clusters ($M_{200}$ ranging from $10^{14.37}\,{\rm M}_\odot$ to $10^{15.79}\,{\rm M}_\odot$, with a median of $10^{15.15}\,{\rm M}_\odot$) compared to the TNG300 sample. 

There are many ways to select galaxies from simulations and hence it is not straightforward to ensure that a fair comparison is being made with the observations.  One reason for this difficulty is the limited realisations of simulations to choose from, and another aspect for our study is that observed protoclusters and their galaxies are detected due to their high SFRs and thus may be biased towards a subset of the population with the highest SFRs (see e.g.\ figure~5 of \citealt{rennehan2020}, which shows a large spread in accretion history from objects at a given $z\,{=}\,0$ mass). What we do know is that the number density of observed protoclusters is estimated to be similar to that of massive clusters with $M_{200}\gtrsim 10^{14.6}\,{\rm M}_\odot$ at $z\,{\simeq}\,0$ \citep{casey2016} and that an average mass-accretion history predicts that most observed protoclusters will evolve into such massive clusters at $z\,{\simeq}\,0$. Given this information, one expects a significant overlap between the high SFR-selected protoclusters and massive clusters with $M_{200}\gtrsim 10^{14.6}\,{\rm M}_\odot$ at $z\,{\simeq}\,0$ (the overlap will not be complete between the two selections because not all structures will follow the average accretion history). This mass limit matches that of the TNG300 sample and is slightly lower than that of the MACSIS sample (as described above), while the number density of both the TNG300 and MACSIS samples is also within the range of the estimated number density of protoclusters from observations. For TNG300, we directly confirm that the samples are similar and thus our results do not change much when a selection is made entirely based on SFR (by rank in SFR, specifically, at each given redshift) instead of this mass cut (see Appendix~\ref{sec_appA}). To further limit selection effects, we present individual simulated clusters for most of our results and subsequent discussions, rather than just the median or average trends. 

We track the growth and accretion history of individual clusters in our sample in order to study their `protocluster phase' by linking them between the simulation snapshots across time. At a given redshift, we define a protocluster by including all galaxies that eventually collapse to a $z\,{=}\,0$ cluster. While this is straightforward for the MACSIS clusters (for which the simulations were run focusing on each of them), it is less clear how to do this for the TNG300 sample and we choose to do this by using the \textsc{SubLink} merger tree to trace them. This is simply done by accounting for all (sub)haloes at a given redshift along the merger trees of all $z\,{=}\,0$ subhaloes. We do this by tracking the accretion history of all individual galaxies and subhaloes using the `LastProgenitorID' and `MainLeafProgenitorID' links, which, when combined, allow us to probe each branch of a $z\,{=}\,0$ subhalo individually within the merger tree of the subhalo. 

The spatial extent of protoclusters defined from observational measurements can be quite ambiguous, and choices for how to do this vary greatly among studies and objects, making it extremely difficult to perform a fair, reliable comparison between objects or with simulations. The spatial extent of the observational sample chosen for our analysis, in particular, spans roughly an order of magnitude, ranging between 0.5 and 5\,Mpc. For consistent analysis of properties across objects from both the simulations and the observations, we use the simulations to convert all properties to those within a common aperture of $5R_{500}$.\footnote{$R_{500}$ is the radius within which the mean enclosed mass density is $500$ times the critical density of the Universe and $M_{500}$ is the mass enclosed within $R_{500}$.} Specifically, we first extract the mean relations between total mass within an observed aperture and $M_{500}$ from the simulations for each observed protocluster of a given redshift and mass (using only simulated protoclusters of similar redshift and aperture mass). We do not find any systematic dependence of the relation on SFR. We then use the obtained relations to estimate $M_{500}$ from a total mass within an observed aperture reported in the original papers for each observed protocluster. The mass conversion uncertainty is found to be significant, typically a factor of $10$, meaning that any study or comparison will fail if mass is not properly calibrated. Using $R_{500}$ computed based on the estimated $M_{500}$, and using the spatial distribution of observed galaxies, we calculate an observed property within $5R_{500}$ by summing up that property for member galaxies within $5R_{500}$. Similarly, we also use the mean relations extracted from the simulations to estimate $M_{200}$ of the observed protoclusters, which is assumed to correspond to that of the `main' halo in the simulated protoclusters. In the simulations, we define the main halo as the FoF halo that the main progenitor\footnote{The main progenitor is the progenitor of a subhalo along its `main progenitor branch', which is defined as the one with the most massive history among all its progenitor branches on the merger tree. The main progenitors are easily traced in time using the `FirstProgenitorID' or `MainLeafProgenitorID' links provided for TNG.} of the most massive subhalo in a cluster at $z\,{=}\,0$ belongs to at a given redshift. While this definition of the main halo has an advantage that it tracks (progenitors of) the same objects across redshifts, thus allowing us to study their evolution, it may be more appropriate to select haloes with the highest SFR at each redshift, given that the observed protoclusters are mainly detected due to their high SFRs. In Appendix~\ref{sec_appA}, we test the SFR-based choice for the definition of the main halo, and demonstrate that this choice does not change the results significantly compared to the much larger discrepancy found with the observations. 

In each FoF halo in the simulations, we refer to the galaxy associated with its most massive subhalo as the `central galaxy', and all the other galaxies as `satellites'. We define the galaxy and cluster `centres' as the positions of the minima of the gravitational potentials that are directly available from both simulations. For each galaxy, we compute the total mass ($M_{\rm tot}$), stellar mass ($M_\ast$), gas mass ($M_{\rm gas}$), and SFR by summing up the relevant quantity for all particles (or cells for the gaseous component in TNG300) of the corresponding type that belong to a given subhalo. Because the gas masses for the observed protoclusters are primarily measured based on low-$J$ CO transitions which trace `cold' gas, we also want to focus on only the cold gas components from the simulations in order for the comparisons to be fair. Unfortunately, there is no simple criterion for defining the cold component in the simulations that perfectly matches that from the observations. As an approximation, we choose to consider only gas cells/particles with $n_{\rm H}\,{>}\,0.1\,{\rm cm}^{-3}$ from the simulations for computing the gas mass. This should be a reasonable approximation because these are particles/cells set to form stars in both simulations (see Sect.~\ref{ssec_TNG} and \ref{ssec_MACSIS}). We use only galaxies with a minimum of $100$ baryonic particles for our analysis, except in Sect.~\ref{sec_spt2349}, where we show all galaxies without a minimum particle limit, but indicate the mass limits corresponding to $100$ particles by plotting appropriate lines. We note, however, that the contribution from the galaxies with less than $100$ particles to the SFRs integrated within $5R_{500}$ is negligible. 

Not all galaxies are detected in given observations due to observational sensitivity. While it is obvious that some criteria for selecting simulated galaxies are needed in order to mimic the observations, a choice for such a criterion is unclear because the sensitivity varies with many factors such as surveyed area, methods used for detection, redshift ranges and properties of individual objects. For simplicity, here we apply a selection cut of ${\rm SFR} \ge 50\,{\rm M_\odot \,yr^{-1}}$, with a Gaussian random scatter of $0.2\,$dex. This choice broadly agrees with the lower limit of SFR that corresponds to a typical detection limit on flux density and line emission for the observations and redshift range considered in our study \citep[e.g.][]{scoville2016, miller2018, hill2020, long2020}. The Gaussian scatter is used to incorporate the dispersion in SFR at a given sensitivity, mainly arising from a scatter in the dust mass, which is also known to affect the flux density \citep[see also][]{hayward2011, hayward2013a, hayward2013b}. As will be seen in Sect.~\ref{sec_spt2349}, the lower limit in gas mass and SFR of the simulated galaxies chosen with this selection is in a good agreement with that of observed galaxies. All results presented in this paper, including the total SFRs of protoclusters, are only based on the simulated galaxies that pass this selection criterion. 

Figure~\ref{fig_TNGsample} shows the gas cells, colour-coded according to their SFRs, associated with the main halo of the protoclusters from the TNG300 sample at a redshift of $4.4$, which is the closest snapshot of TNG300 to the redshift of the SPT2349$-$56 structure. In each panel, the largest circle shows the location of $R_{500}$ for the main halo, while the smaller circles indicate subhaloes. One can see that there is a great amount of variety in the morphology of the simulated protoclusters.

%%%%%%%%%%% figure
\begin{figure*}
\includegraphics[width=0.95\linewidth]{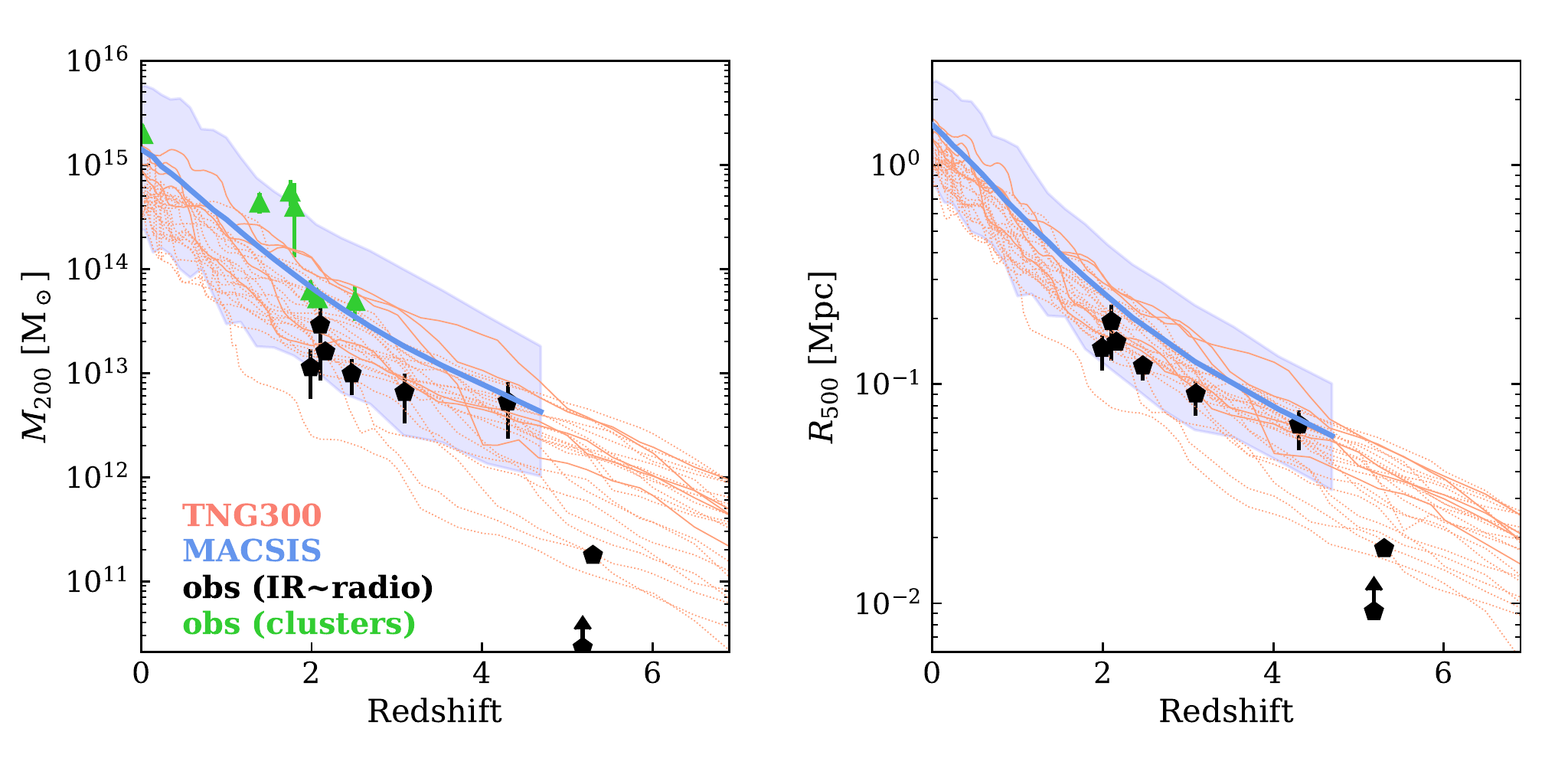}
\caption{Evolution of the main halo mass (left panel) and size (right panel) of the 25 most massive $z=0$ clusters from TNG300 (salmon lines) to $z\,{\simeq}\, 7$, together with the median and 99.5 per cent range from the entire MACSIS sample (blue line and band, respectively). For the TNG300 results, the solid lines are used for a subset consisting of the five most massive clusters, while the dotted lines are for the rest of the sample. The pentagons represent an observational compilation from the literature, as described in Sect.~\ref{sec_obs}, with the correction for aperture mass described in Sect.~\ref{sec_methods}; this consists of the seven protoclusters/overdensities from \citet{casey2016} and SPT2349$-$56 at $z\,{=}\, 4.3$ from \citet{hill2020}. The green triangles show massive clusters at lower redshift, namely XMMU J2235.3 \citep{rosati2009}, XLSSC 122 \citep{mantz2018}, IDCS J1426.5$+$3508 \citep{stanford2012, brodwin2016}, JKCS 041 \citep{andreon2014}, CL J1449$+$0856 \citep{gobat2011, gobat2013} and CL J1001$+$0220 \citep{wang2016}, as well as the Coma cluster, for a comparison of their evolution over a wider range of redshifts. We find that the mass evolution of the simulated clusters is described well, within a typical scatter of $\simeq 0.2\,$dex, by the fitting function given in Eq.~\ref{eq_MAH}. }
\label{fig_MAH}
\end{figure*}

%%%%%%%%%%% figure
\begin{figure*}
\includegraphics[width=1.0\linewidth]{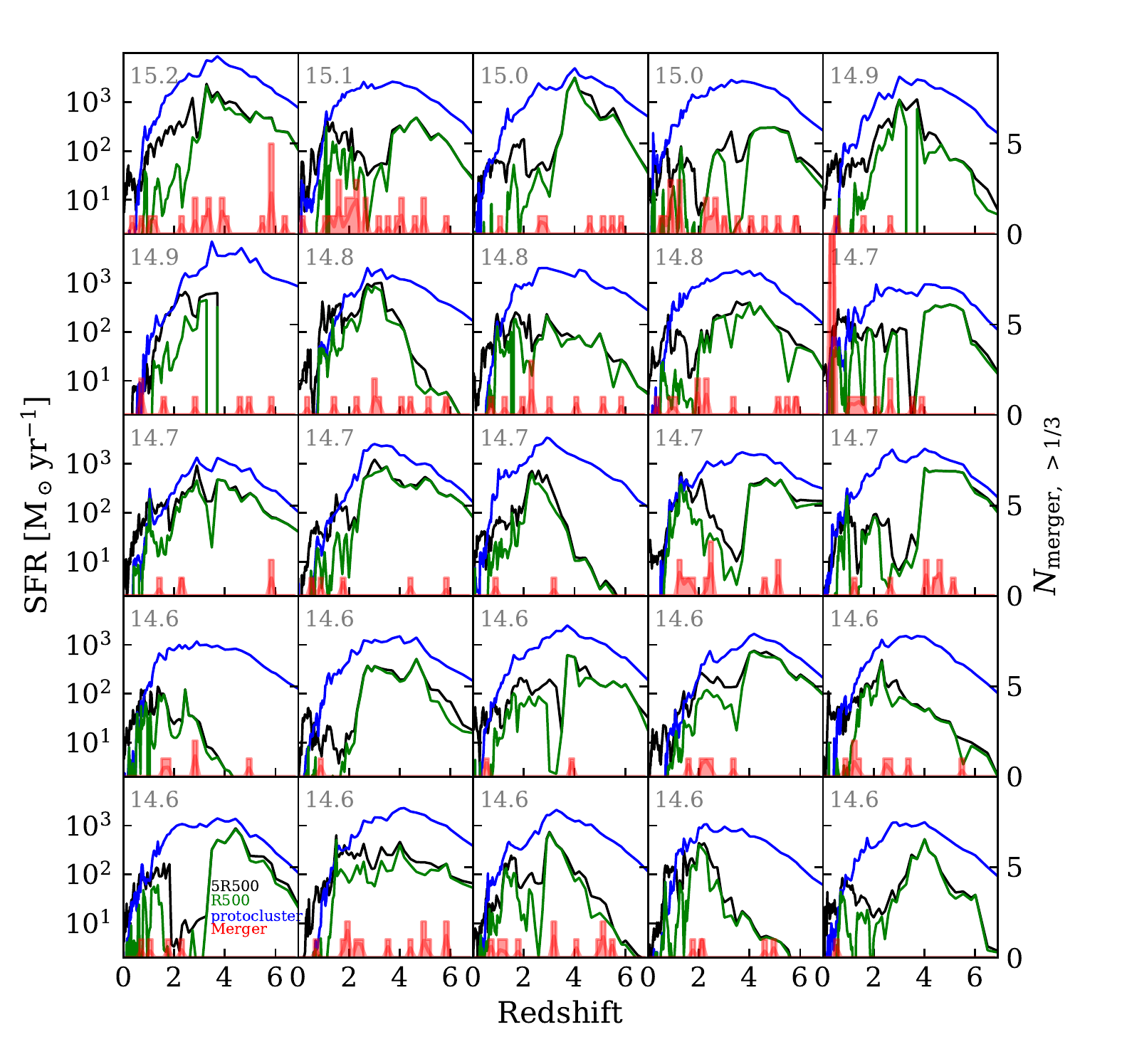}
\caption{Star-formation history of the 25 most massive $z=0$ clusters from TNG300, calculated within $5R_{500}$ (black) and $R_{500}$ (green), centred on their main halo, and for the entire protocluster (blue). The $M_{200}$ value of each cluster at $z=0$ is indicated in each panel as a logarithm in solar mass units. The red histograms show the number of mergers with a mass ratio greater than 3:1 that occurred with the main halo in a given redshift bin, while the red lines indicate the same number of mergers but weighted with the mass ratio of the mergers. }
\label{fig_SFH_TNG}
\end{figure*}

%%%%%%% SECTION 5
\section[results]{RESULTS}
\label{sec_results}

\subsection{Mass-accretion history}

One of the key properties that describes a halo is its mass: halo mass is known to correlate (weakly) with many of the other halo properties, such as size, formation time, concentration, spin parameter, clustering and environment \citep[e.g.][]{fall1980, mo1996, sheth2001, zhao2003, zhao2009}, although with a large spread \citep[e.g.][]{rennehan2020}. Because galaxies are believed to form and evolve within their dark matter haloes, the halo mass is further known to have an important impact on galaxy evolution. Thus a halo's mass also correlates with properties of the galaxies embedded within it, including galaxy number density, size, luminosity, morphology, angular momentum, colour, stellar age, SFR and gas contents \citep[e.g.][]{vogelsberger14b, schaye2015, mccarthy2017, lim18, pillepich18b, wang2018, behroozi2019}. 

Comparing the masses of the observed protoclusters with those of the simulated clusters, therefore, will not only confirm whether the simulations contain an appropriate population that corresponds to the observational samples, but will also help to determine approximately the evolution of the protoclusters (such as their mass at $z\,{=}\,0$, which is not directly available from observations). In Fig.~\ref{fig_MAH} we investigate the MAH of the TNG300 and MACSIS cluster samples out to $z\,{\simeq}\,7$, as described in Sect.~\ref{sec_methods}, and compare them with that from the observational data (symbols), as compiled in Sect.~\ref{sec_obs} and corrected for mass in Sect.~\ref{sec_methods}. For comparison, we also plot some massive clusters at lower redshift with mass derived mainly from X-ray observation, shown by the green triangles, which include XMMU J2235.3 \citep{rosati2009}, XLSSC 122 \citep{mantz2018}, IDCS J1426.5$+$3508 \citep{stanford2012, brodwin2016}, JKCS 041 \citep{andreon2014}, CL J1449$+$0856 \citep{gobat2011, gobat2013} \citep[see also][for mass]{valentino16} and CL J1001$+$0220 \citep{wang2016}, as well as the Coma cluster. It is worth noting that the mass for the observational sample from the original papers is not always estimated in the same way; for some of them the total mass is obtained via abundance matching, based on the estimated stellar masses, which are derived from luminosities with an assumed mass-to-light ratio, while for others it is estimated based on the galaxy overdensity or dynamical mass. 

As can be seen, $M_{200}$ for the observational sample is comparable to that of the TNG300 sample, meaning that the TNG300 cluster samples are fairly representative of the observed (proto)clusters, regarding their masses. The MACSIS sample is about 2--3 times more massive on average than the TNG300 sample at any given redshift. This is because the average mass of the sample is higher for MACSIS than for TNG300 by about the same factor (see Sect.~\ref{sec_methods}). We also use the solid lines to indicate the results from a subset consisting of the five most massive (at $z\,{=}\,0$) clusters from the TNG300 sample, while using the dotted lines to show the results from the rest of the sample. The subset of the TNG300 sample is consistent with the MACSIS sample regarding the MAH. The mass estimates for a few protoclusters observed at around $z\,{\simeq}\,5$ (HDF\,850.1 and AzTEC-3 structure) are found to be much lower than the average MAH from the simulations. While this can be a result of stochasticity in the MAH for individual clusters (and a few of the TNG300 clusters are indeed found to have masses consistent with these observed protoclusters in the same redshift regime, but still become clusters with $\simeq 10^{15}\,{\rm M}_\odot$ at $z\,{=}\,0$), it could also be due to the mass estimates from observations at the highest redshifts being highly uncertain. We find that the MAHs of the simulated clusters are described well by the following fitting function, with a typical scatter of about $0.2\,$dex: 
%%%%%%%%%%%%%%%%%%%%%%%%%%%%%%%%%%%
\begin{eqnarray}\label{eq_MAH}
\log_{10}\Big(\frac{M_{200}(z)}{M_{200}(z=0)}\Big) = -0.741z + 0.0322z^2 - 0.00237z^3. 
\end{eqnarray}
%%%%%%%%%%%%%%%%%%%%%%%%%%%%%%%%%%%

In the right panel of Fig.~\ref{fig_MAH} we also show the size evolution of the simulated clusters, compared with $R_{500}$ of the observed protoclusters estimated in Sect.~\ref{sec_methods}. The $R_{500}$ value estimated for the SPT2349$-$56 protocluster, for example, is remarkably consistent with the observational estimate of between 65 and 90\,kpc \citep[][hereafter H20]{hill2020}, and also consistent with the range of $R_{500}$ values from the simulations at the same redshift of $z\,{\simeq}\, 4.3$.

\subsection{Star-formation history}

Despite the similarity in MAHs, this does not tell us much about the gaseous and stellar components, since the total mass is dominated by the dark matter rather than the baryons, and thus is mostly governed by gravity and cosmology, rather than by galaxy-scale physics. A more detailed and quantitative comparison with observations can thus be made through a direct examination of the properties of the baryons. 

One of the key properties of galaxies is the SFR, since it is shown to have strong correlations with many other galaxy properties, such as colour, age and morphology. Figure~\ref{fig_SFH_TNG} investigates the star-formation history (SFH) of the 25 massive clusters from TNG300 out to $z\,{\simeq}\,7$. We specifically show the SFH of protoclusters, i.e. all galaxies and subhaloes that become a member of a given cluster at $z\,{=}\,0$ (as defined in Sect.~\ref{sec_methods}), together with the SFH calculated within $5R_{500}$ and $R_{500}$ of the main halo (also defined in Sect.~\ref{sec_methods}). One can see that most of the star formation in the protoclusters at high redshift occurs outside the main halo, while star formation in the main halo is relatively centralized, with only a small fraction of stars forming outside its $R_{500}$ radius on average (although the scatter between the individual clusters and for different redshifts is large). At low redshift, the trends are reversed, so that most star formation occurs within the main halo but outside its $R_{500}$ radius. Note that the integrated SFR shown here includes the contribution from galaxies that are not associated with the main halo. Although we find that such contributions are insignificant at $z\,{\gtrsim}\,2$, which our analysis focuses on, the contribution is significant at the lower redshifts, such that the SFH of (proto)clusters (blue curve) is lower than that within $5R_{500}$ (black curve) at $z\,{\lesssim}\,1$.

%%%%%%%%%%% figure
\begin{figure*}
\includegraphics[width=0.85\linewidth]{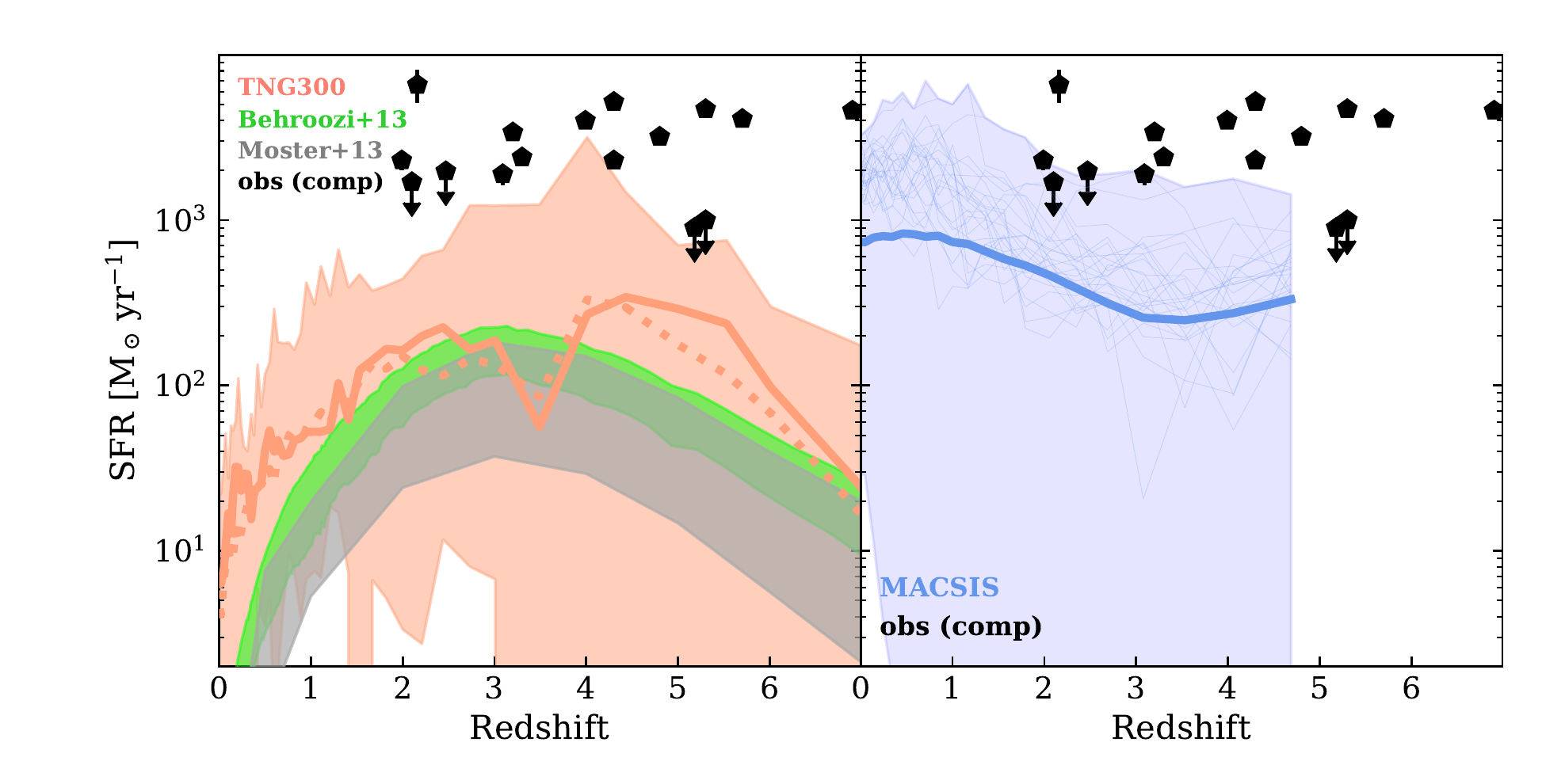}
\caption{Star-formation history, calculated within $5R_{500}$, of the main halo for the 25 most massive $z=0$ clusters from TNG300 (salmon line and band showing the median and 99.5 per cent range, respectively), and of the entire MACSIS sample (blue line and band showing the median and 99.5 per cent range, respectively). For the TNG300 results, the solid line is the median SFH for a subset consisting of the five most massive clusters, while the dotted line is the median from the whole sample. We also show the results for the 25 most massive $z=0$ clusters from the MACSIS sample as the thin lines in the right panel. The symbols show a compilation of observations from the literature, as described in Sect.~\ref{sec_obs} and corrected to the aperture of $5R_{500}$ in Sect.~\ref{sec_methods}: the seven protoclusters/overdensities from \citet{casey2016}; the eight SPT protocluster candidates from \citet{wang2020}; and SPT2349$-$56 from \citet{hill2020}. The predictions from the empirical models of \citet{behroozi2013} (green) and \citet{moster2013} (grey) are shown for comparison. }
\label{fig_SFH}
\end{figure*}

%%%%%%%%%%% figure
\begin{figure}
\includegraphics[width=0.93\linewidth]{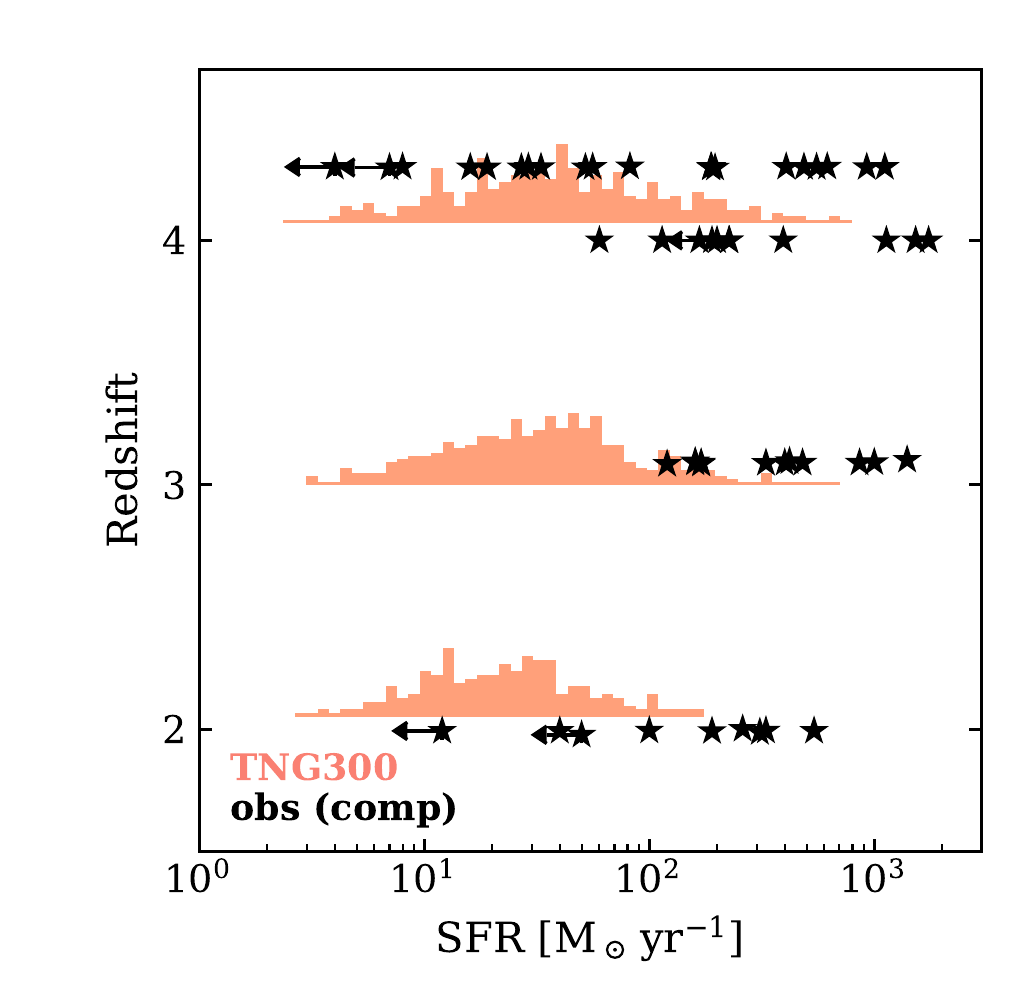}
\caption{Histogram (with arbitrary normalizations) of the member galaxy SFRs of the 25 most massive protoclusters from TNG300, shown specifically at $z\,{\simeq}\,2$, $3$, and $4$ (salmon). The symbols show the observational compilation, which includes the GOODS-N protocluster at $z\,{=}\,2.0$, the SSA22 structure at $z\,{=}\,3.1$ (both with their SFR estimates adopted from \citealt{casey2016}), the DRC at $z\,{=}\,4.0$ from \citet{long2020} and SPT2349$-$56 at $z\,{=}\,4.3$ with the SFR estimated by \citet{hill2020}. }
\label{fig_SFR}
\end{figure}

Figure~\ref{fig_SFH} compares the median and 99.5 per cent range of the SFH of the TNG300 and MACSIS samples with the observations and with the model predictions of B13 and M13. We also show, using thin lines, the SFHs of the 25 most massive clusters from the MACSIS sample. The median SFH from the subset consisting of the five most massive clusters in the TNG300 sample is consistent with that from the whole sample, except at $z\,{\gtrsim}\, 4.5$, where the former is higher by a factor of approximately 2 compared to the latter. Both the simulations and the empirical models predict much lower SFRs than that seen in the observations, with the upper limit from the simulations marginally overlapping with the lower limit from the observations. Given the similarity in the MAH between the observations and the simulations previously shown in Fig.~\ref{fig_MAH}, the large discrepancy seen here indicates that the SFRs predicted by the simulations are much lower than the observations at a given mass. The discrepancy with observation by more than a factor of 2--4 shown here is consistent with findings of some previous studies \citep[e.g.][]{granato15, bassini20} that used completely different sets of simulations and subgrid models,  indicating that the issue lies not with the particular simulations and models, but is more general. The predictions of the empirical models are even lower than the simulations, typically by a factor of 2--3. Note that the uncertainty shown for the model of B13 is the standard error of the mean; the scatter between individual clusters predicted by B13 is not available because of the mass limit of the merger tree used by B13, but it is expected to be much larger than the standard error in the mean, by up to 1--1.5\,dex \citep{behroozi2013}. In addition, the MACSIS simulations predict much higher SFRs than TNG300, particularly at low redshift. However, the discrepancy between the two simulations is shown to be diminished when only those with the highest SFRs are compared, particularly at $z\,{\gtrsim}\,3$, implying that part of this is due to selection (see Appendix~\ref{sec_appA}). Finally, we find that the galaxies lying below the selection criteria described in Sect.~\ref{sec_methods} contribute about 15 per cent of the total SFR at any given redshift, which is relatively negligible. 

In Fig.~\ref{fig_SFR}, we also compare the SFRs of individual member galaxies of the protoclusters at high redshift. Specifically we choose four high-$z$ protoclusters from the observational compilation that have SFR estimates for each individual galaxy, which are the GOODS-N protocluster at $z\,{=}\,1.99$, the SSA22 structure at $z\,{=}\,3.09$ (both with their SFR estimates adopted from \citealt{casey2016}), the DRC at $z\,{=}\,4.00$ from \citet{long2020} and SPT2349$-$56 at $z\,{=}\,4.3$ (with the SFR estimated by H20). The SFR for a given galaxy in the simulations is calculated by summing up the SFR of all particles/cells that are associated with a given subhalo. We find that using different apertures for summing up the SFR does not change our conclusions. Note that here we plot all progenitors from the simulation belonging to the TNG300 clusters at $z\,{=}\,0$, not just those in the main haloes. It can be seen not only that the observed protocluster galaxies have much higher SFRs on average, but that there are multiple galaxies observed with SFRs higher than $\simeq 10^3\,{\rm M_\odot\,yr^{-1}}$ even in a single protocluster; these are not predicted at all from the simulations.

\subsection{Concentration of star formation}

\begin{figure*}
\includegraphics[width=1.0\linewidth]{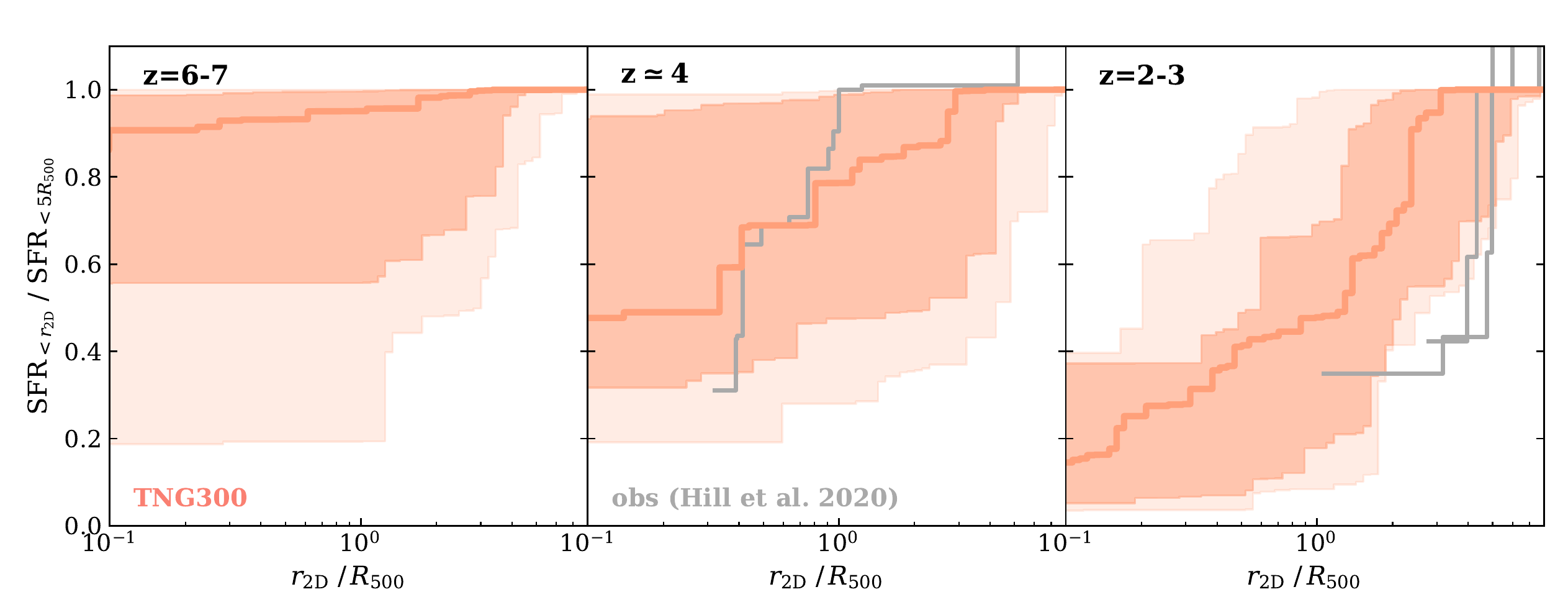}
\caption{Cumulative SFR profile, normalized by that at $5R_{500}$, for the main haloes of the 25 most massive protoclusters from TNG300 at redshifts of $6$--$7$ (left panel), $4.4$ (middle panel) and $2.2$ (right panel), with the salmon lines and bands showing the median, 68\,per cent and 99.5\,per cent ranges, respectively. The grey step functions show the observational compilation, with the SFR profiles for SPT2349$-$56 (middle; from \citet{hill2020}), and the SSA22 structure, the GOODS-N protocluster and MRC1138$-$262 (right; all from \citet{hill2020}). The $R_{500}$ values are estimated based on \citet{hill2020} (middle) and \citet{casey2016}, using the correction to aperture mass as described in Sect.~\ref{sec_methods}. }
\label{fig_prof}
\end{figure*}

We also investigate the concentration of the star formation seen within protoclusters by examining the cumulative fraction of SFR within apertures. Such SFR profiles may give us hints for where the discrepancy in the total SFR between observations and simulations might come from. Figure~\ref{fig_prof} shows the integrated SFR within apertures of given radii with respect to the total SFR within $5R_{500}$, as predicted for the TNG300 sample. They are compared with those from the observations, using the SFR profiles adopted from H20. The scatter between the simulated clusters is large, particularly at the higher redshifts of $z\,{\gtrsim}\,4$, with the SFR fraction ranging from $0.2$ to nearly $1$ at $R_{500}$, implying a great diversity within the high-$z$ protocluster population, despite their similarity in mass at $z\,{=}\,0$. The profiles from TNG300 are systematically higher on average in the inner regions (i.e. star formation is more centralized) than seen in the observations, although the difference is within the scatter. The increasing concentration of SFR with increasing redshift, as predicted from the simulation, is consistent with the observations, although the change occurs more rapidly for the observed protoclusters. We find that the selection criteria for the simulated galaxies (as described in Sect.~\ref{sec_methods}) has a negligible impact on the profile, implying that the observed profiles are also likely insensitive to the current detection limit. Additionally, the results from TNG300-3 or MACSIS are broadly consistent with those from TNG300-1, within the uncertainties, indicating that the profile does not depend strongly on numerical resolution or an overall amplitude of the SFR. We also confirm that the median profiles from the massive subset of the TNG300 sample are consistent with those from the whole sample; thus the profile is insensitive to mass for the mass range of the sample. 

%%%%%%%%%%% figure

%%%%%%% SECTION 6
\section[SPT2349]{SPT2349$-$56 as a case study}
\label{sec_spt2349}

In this section, we present a more thorough analysis of the SPT2349$-$56 protocluster as a case study to help improve our understanding of the modelling of protoclusters at high redshift and their comparison to observations. As described in Sect.~\ref{sec_obs}, SPT2349$-$56 is one of the highest-redshift protoclusters identified so far, located at $z\,{=}\, 4.3$; it shows an extremely high SFR in the central region of the structure, which consists of more than 20 galaxies, about half a dozen of which have SFRs greater than $400\,{\rm M_\odot\,yr^{-1}}$ within a projected radius of $90\,$kpc (see H20). Here we explicitly consider 23 member galaxies in the densest part of this structure, as recently identified and studied in H20, with estimates of their properties also adopted from the same paper. Note that the protocluster has two more regions with clustered star formation (but with fewer galaxies identified) located farther from the centre (at about $400\,$kpc and $2\,$Mpc in projection), which we do not include in our analysis. The central region of SPT2349$-$56 has a size of about $90$\,kpc in radius, and an analysis of the velocity distribution of the galaxies suggests that the region is more or less virialized, with an estimated mass of $(9\pm 5) \times 10^{12}\,{\rm M_\odot}$ (H20). The integrated SFR over all three regions is estimated to be upwards of $10,000\,{\rm M_\odot\,yr^{-1}}$, and the comoving volume encompassed by the entire structure is a few thousand ${\rm cMpc}^3$. The structure is expected to evolve to become at least a Coma-cluster-like object in mass at $z\,{\simeq}\,0$.

\subsection{Star-formation rates}
\label{ssec_SFR}

%%%%%%%%%%% figure
\begin{figure*}
\includegraphics[width=1.0\linewidth]{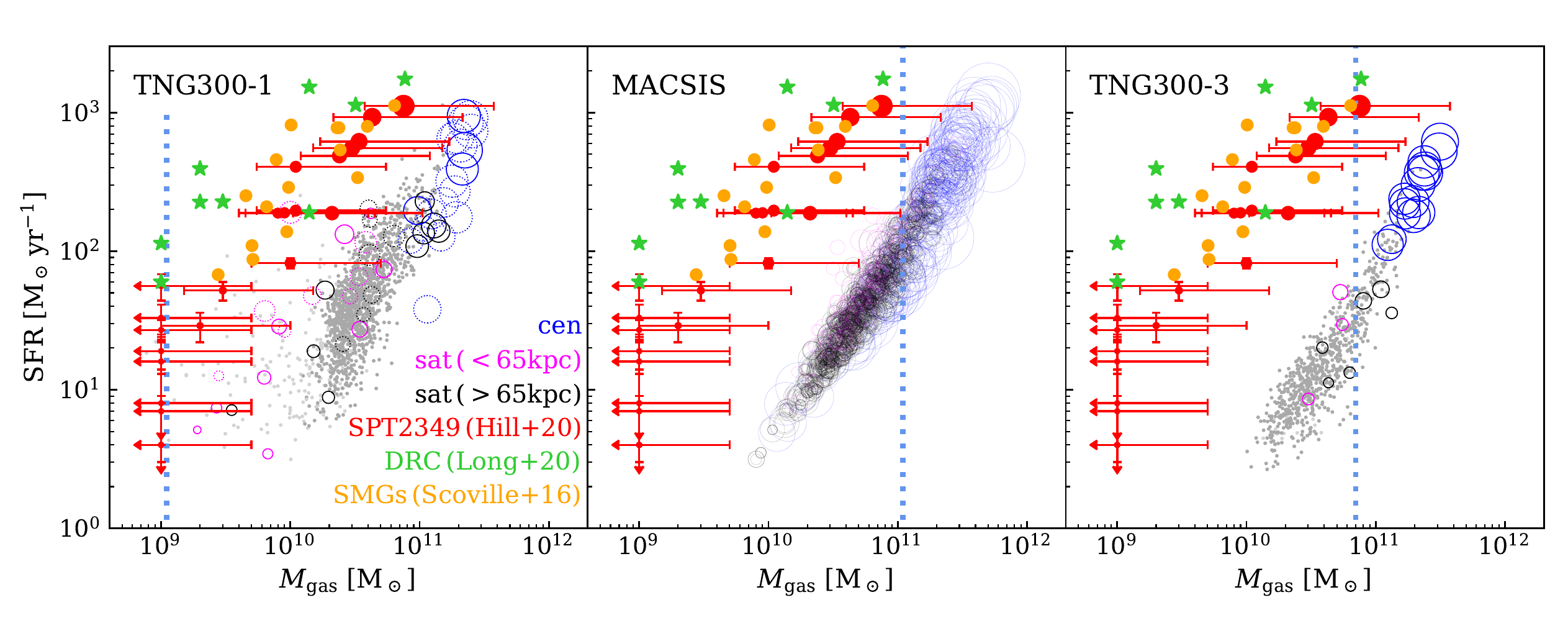}
\caption{SFRs of galaxies in the main halo of the simulated protoclusters (unfilled circles) at $z\,{\simeq}\, 4.3$, shown separately for centrals (blue) and for satellites within (magenta) and outside (black) $65\,$kpc (which is the estimated $R_{500}$ value for SPT2349$-$56 from Sect.~\ref{sec_results}) from the centre. The results are shown for TNG300-1 (left), MACSIS(middle) and TNG300-3 (right). For TNG300-1, the galaxies belonging to the five most massive clusters are shown by the solid circles, while the dotted circles indicate the others. The dark (light) grey dots show other central (satellite) galaxies from the simulations that do not belong to the sampled protoclusters. The vertical lines indicate the mass limits corresponding to the cut of 100 particles for each simulation (Sect.~\ref{sec_methods}). For comparison, the red circles show the observations for the member galaxies of the SPT2349$-$56 protocluster from \citet{hill2020}, while the green stars show the DRC galaxies based on \citet{long2020}, but with the gas mass estimated directly from CO luminosity. The sizes of the circles for both the simulations and SPT2349$-$56 are proportional to the sizes of the subhaloes associated with the galaxies, calculated based on the total mass of subhaloes, with the correction for mass loss due to stripping as described in the text. Finally, the orange dots show SMGs at $z\,{=}\,4$--6 observed with ALMA, from \citet{scoville2016}. }
\label{fig_SFR_Mgas}
\end{figure*}

We begin by examining the SFRs of member galaxies as a function of their gas mass, $M_{\rm gas}$, i.e. the star-formation efficiency, which is shown in Fig.~\ref{fig_SFR_Mgas}. The SFRs and gas masses of the SPT2349$-$56 galaxies are adopted from H20, where they were estimated using their far-infrared (FIR) luminosities and CO line-emission strengths, respectively. For comparison, we also present the same properties for the DRC galaxies, which are at a similar redshift ($z\,{\simeq}\,4.0$), based on \citet{long2020}. Among the total of 11 galaxies from the original paper, we exclude those labelled DRC-4 and DRC-5 because of the large uncertainties in their measurements. While the SFRs for the DRC galaxies were obtained using SED fitting in a similar way as with SPT2349$-$56 in H20, we find \citep[as also pointed out by][]{long2020} that their gas mass estimates that are based on a calibration of the dust continuum are about an order of magnitude higher than those that are based on CO luminosity. While it is not entirely clear which method is more reliable, we choose to estimate the gas mass based on CO luminosity and use that procedure in our comparisons, for the sake of consistency with SPT2349$-$56. It can be seen that the SFR and star-formation efficiency (as well as the gas mass distribution) of SPT2349$-$56 broadly matches that of the DRC, given the large uncertainties in SFR estimates for the DRC galaxies (with a typical uncertainty comparable to the signal). 

In the left panel of Fig.~\ref{fig_SFR_Mgas}, the observations are compared with galaxies within $5R_{500}$ of the main halo of the TNG300 sample at its $z\,{\simeq}\,4.4$ snapshot, which is the closest to the redshift of SPT2349$-$56. As described in Sect.~\ref{sec_methods}, we only select and plot the galaxies that pass the cut of ${\rm SFR} \ge 50\,{\rm M_\odot \,yr^{-1}}$, with a Gaussian scatter of $0.2\,$dex to crudely mimic the observational selections. For the simulated galaxies, by using different colours, we separate in the figure the centrals (blue) and satellites within (magenta) and outside (black) a projected distance of $65\,$kpc from the centre. The value $65\,$kpc is the estimated $R_{500}$ for SPT2349$-$56 from Sect.~\ref{sec_results} and also matches the size of the `core' of the structure defined in \citet{miller2018}. The sizes of the circles in the figure for both the SPT2349$-$56 and simulated galaxies is proportional to the size of the subhaloes that each individual galaxy is associated with, which is derived from their total masses (including the dark-matter mass). The total masses of satellite galaxies are estimated assuming that they follow the same gas-to-total mass ratio as the central galaxies. This procedure thus accounts for mass loss due to stripping for satellites after their accretion onto a larger halo. We apply the same gas-to-total mass relation from central galaxies in the simulations, to estimate the total mass for the SPT2349$-$56 galaxies. The total mass summed over all SPT2349$-$56 galaxies estimated in this way is ${\simeq}\,5.2\times 10^{12}\,{\rm M_\odot}$, broadly consistent with the dynamical mass estimate of the entire structure, $(9\pm 5) \times 10^{12}\,{\rm M_\odot}$ from \cite{hill2020}.  However, note that the dynamical mass takes into account all contributions to mass within the structure, and is thus supposed to be higher than a mass sum of individual member galaxies as computed by the model. 

It can clearly be seen in the left panel of Fig.~\ref{fig_SFR_Mgas} that the SFRs are more than an order of magnitude higher at a given gas mass for the SPT2349$-$56 galaxies than for the simulated ones, over a span of a factor of $100$ in gas mass. The uncertainty in the observational measurements, indicated by the error bars, is dominated by the CO-to-gas conversion factor, $\alpha_{\rm CO}$, which ranges typically between 0.5 and 5 in the literature. For the data points, we fix the conversion factor to the value adopted in H20, which is exactly 1. Another interesting discrepancy found between the observations and the simulations is that SPT2349$-$56 is apparently missing a distinct `central' galaxy; SPT2349$-$56 lacks a clear separation between its central galaxy and satellites, unlike for the protoclusters in TNG300 that have centrals that clearly stand out because of their large gas reservoirs (with $M_{\rm gas}\simeq 10^{12}\,{\rm M_\odot}$), but also because of their much higher SFRs and total masses (as indicated in the figure by the sizes of the circles) than for satellites.  The most massive member of the SPT2349$-$56 system only has its gas mass and total mass comparable to those of massive satellites from the simulations, while their SFRs are all comparable to that of centrals from the simulations. This may indicate that SPT2349$-$56 is at an earlier stage of evolution than a typical protocluster in the simulations. 

The middle panel of Fig.~\ref{fig_SFR_Mgas} presents the same comparison, but now for the MACSIS protoclusters at its $z\,{=}\,4.1$ snapshot, the snapshot closest to SPT2349$-$56 in redshift. One can see that the discrepancy in SFR relative to the SPT2349$-$56 galaxies is similar to that from TNG300 at a high gas mass of $M_{\rm gas}\gtrsim 10^{11}\,{\rm M_\odot}$, but becomes more significant at lower gas mass. We discuss this in more detail in Sect.~\ref{sec_discussion}, focusing on the issue of numerical convergence. 

For both the TNG300 and MACSIS samples we also see that the simulated galaxies within $65\,$kpc have higher SFRs on average than those outside of this radius at a given gas mass and total mass. This indicates that galaxies in denser regions have more triggered star formation, which is consistent with a model where tidal interactions occurring in mergers or dense environments can elevate star formation. We do not see any difference between the galaxies from the massive subset of the TNG300 sample (circles) and from the whole sample (dots). We find that using gas mass or SFR integrated in a different aperture for the simulated galaxies does not change any of the conclusions qualitatively, mainly because a change in aperture only moves the galaxies diagonally along the SFR--gas-mass sequence. Clearly, both TNG300 and MACSIS simulations predict much less active star formation than what is found among the SPT2349$-$56 galaxies of similar gas mass. In Sect.~\ref{sec_discussion} we discuss possible origins and explanations for these discrepancies, as well as limitations in making the comparisons.

\subsection{Gas fraction}
\label{ssec_fgas_fb}

%%%%%%%%%%% figure
\begin{figure*}
\includegraphics[width=1.0\linewidth]{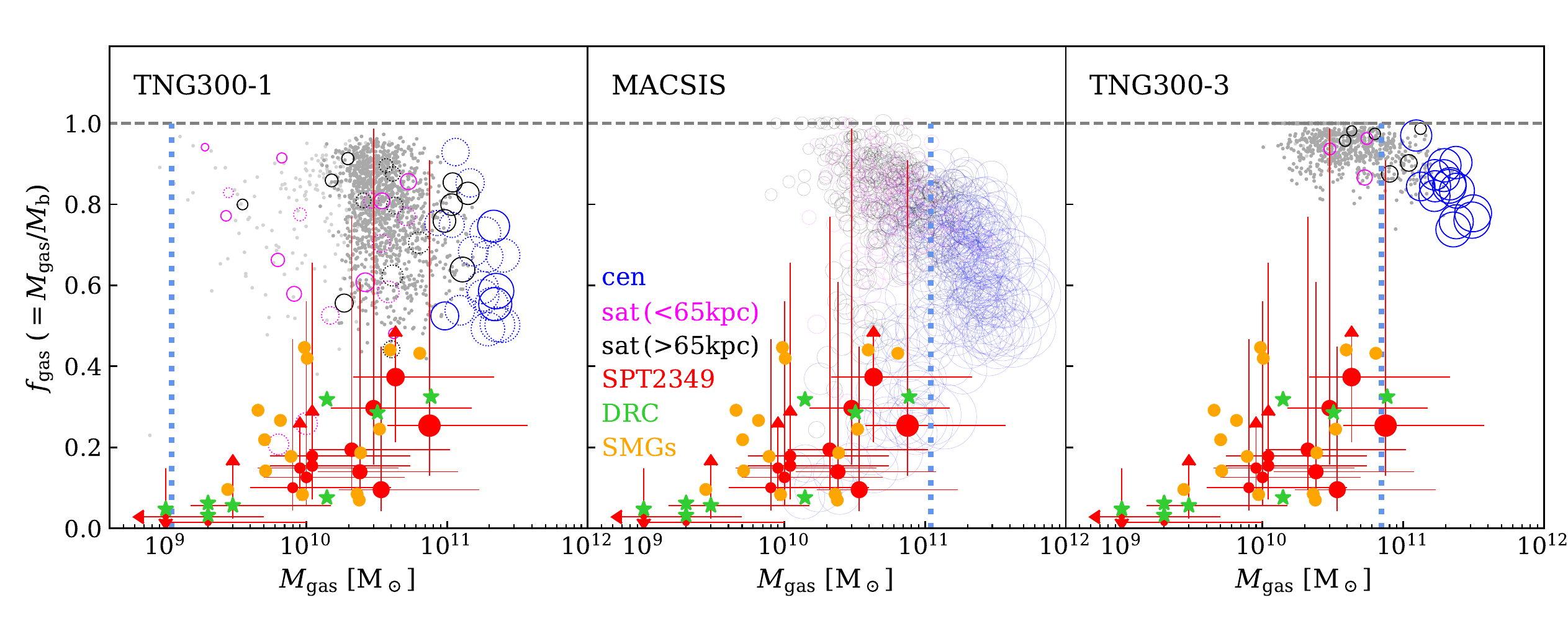}
\caption{Gas fractions of galaxies in the main halo of the simulated protoclusters (unfilled circles) at $z\,{\simeq}\, 4.3$, shown separately for centrals (blue) and for satellites within (magenta) and outside (black) $65\,$kpc from the centre. The results are shown for TNG300-1 (left), MACSIS (middle) and TNG300-3 (right). For TNG300-1, the galaxies belonging to the five most massive clusters are shown by the solid circles, while the dotted circles indicate the others. The dark (light) grey dots show other central (satellite) galaxies from the simulations that do not belong to the sampled protoclusters. The vertical lines indicate the mass limits corresponding to the limit of 100 particles for the definition of objects in each simulation (Sect.~\ref{sec_methods}). For comparison, the red circles show the observations for the member galaxies of the SPT2349$-$56 protocluster, with their gas masses and stellar masses adopted from \citet{hill2020} and \citet{Rotermund}, respectively. The green stars show the DRC galaxies based on \citet{long2020}, but with the gas mass estimated directly from CO luminosity. The sizes of the circles for both the simulations and SPT2349$-$56 are proportional to the sizes of the subhaloes associated with the galaxies, calculated based on the total mass of subhaloes, with the correction for mass loss due to stripping as described in Sect.~\ref{ssec_SFR}. Finally, the orange dots show the SMGs at $z\,{=}\,4$--6 observed with ALMA, from \citet{scoville2016}. } 
\label{fig_fgas}
\end{figure*}

To further explore the characteristics of galaxies in protoclusters, in this subsection we compare the gas fractions of the SPT2349$-$56 galaxies with those of the simulated galaxies. We adopt the stellar mass estimates for the SPT2349$-$56 galaxies from \citet{Rotermund}, which are estimated using SED fitting. Figure~\ref{fig_fgas} compares the gas mass fraction of each individual galaxy from the simulations with the estimates for SPT2349$-$56. The samples shown and the meanings of the colours and sizes of the circles are the same as in Fig.~\ref{fig_SFR_Mgas}. For comparison we show the stellar masses for the DRC galaxies taken from \citet{long2020}, estimated from SED fitting. As was the case for the SFR-gas mass comparison, the DRC galaxies are located very close to the SPT2349$-$56 galaxies, indicating that they belong to a similar population at their redshift. The simulated galaxies, on the other hand, show a significantly higher gas-mass fraction of $\simeq0.7$ and above. This is likely due to their much lower SFRs at a given gas mass, as shown previously, leading to smaller numbers of stars forming relative to the observations. Estimates of $f_{\rm gas}$ obtained from observations at high redshifts typically show a wide range of values: for instance, \citet{tadaki2019} measured a median $f_{\rm gas}$ of $0.77$, with a spread from $0.4$ to $0.9$, for galaxies in three protoclusters at $z=2.16$--$2.53$ based on their ALMA CO(3--2) observations, which is more or less consistent with the predicted values from the simulations. The gas fraction for SPT2349$-$56 is, on the other hand, significantly lower, with an average value of $\simeq 0.2$, although the measurement error is large due to the uncertainty in the conversion factors. Also note that the previously mentioned discrepancy by a factor of 10 between the gas mass estimated from CO luminosity versus  dust continuum can by itself elevate the average $f_{\rm gas}$ of SPT2349$-$56 to around 0.7.

\subsection{Comparison with field SMGs}

One of the findings from \citet{Rotermund} is that the SPT2349$-$56 galaxies do not appear to stand out relative to field SMGs, i.e. galaxies that are identified with high SFR, but do not belong to observationally identified protoclusters. We use the words `field SMGs' here even although they are not necessarily field galaxies in a strict sense, since they might still belong to a dense environment or be involved in mergers (as suggested by their high SFRs). We take the high-$z$ SMG samples from \citet{scoville2016}, which were observed with ALMA at redshifts between 4 and 6, as field SMGs to be compared with the protocluster SMGs. The gas masses of the field SMGs in the original paper were estimated using an approach similar to that of \citet{long2020}, which leads to higher gas mass estimates by around an order of magnitude compared with estimates based on CO luminosity. We thus scale the gas mass estimates by 1/10 for a consistent comparison between the observations. 

In Figs.~\ref{fig_SFR_Mgas} and \ref{fig_fgas}, the field SMGs chosen are compared with the protocluster galaxies from both observations and simulations. From the simulations, we apply the same criteria, i.e. any galaxy with a high SFR that does not belong to a protocluster with a mass similar to the observed protoclusters is defined and selected as a `field' SMG. From the observations, the field SMGs have properties that are remarkably similar to those of the protocluster galaxies. This indicates that the observed protoclusters, even including SPT2349$-$56, stand out only because of their assembly, i.e. by the extreme overdensity of SMGs. However, in the simulations, the protocluster galaxies clearly show higher SFRs (and thus higher star-formation efficiencies) and lower gas-mass fractions than field galaxies at a given gas mass. This is an indication that, in the simulations, star formation in galaxies is induced and boosted more by mergers and interactions in denser environments, which was also indicated by the higher SFRs in galaxies found in the inner regions compared to the outer regions of the simulated protoclusters, unlike in the observations.

\subsection{Abundance and mass function}
\label{ssec_gmf}

\begin{figure}
\includegraphics[width=1.0\linewidth]{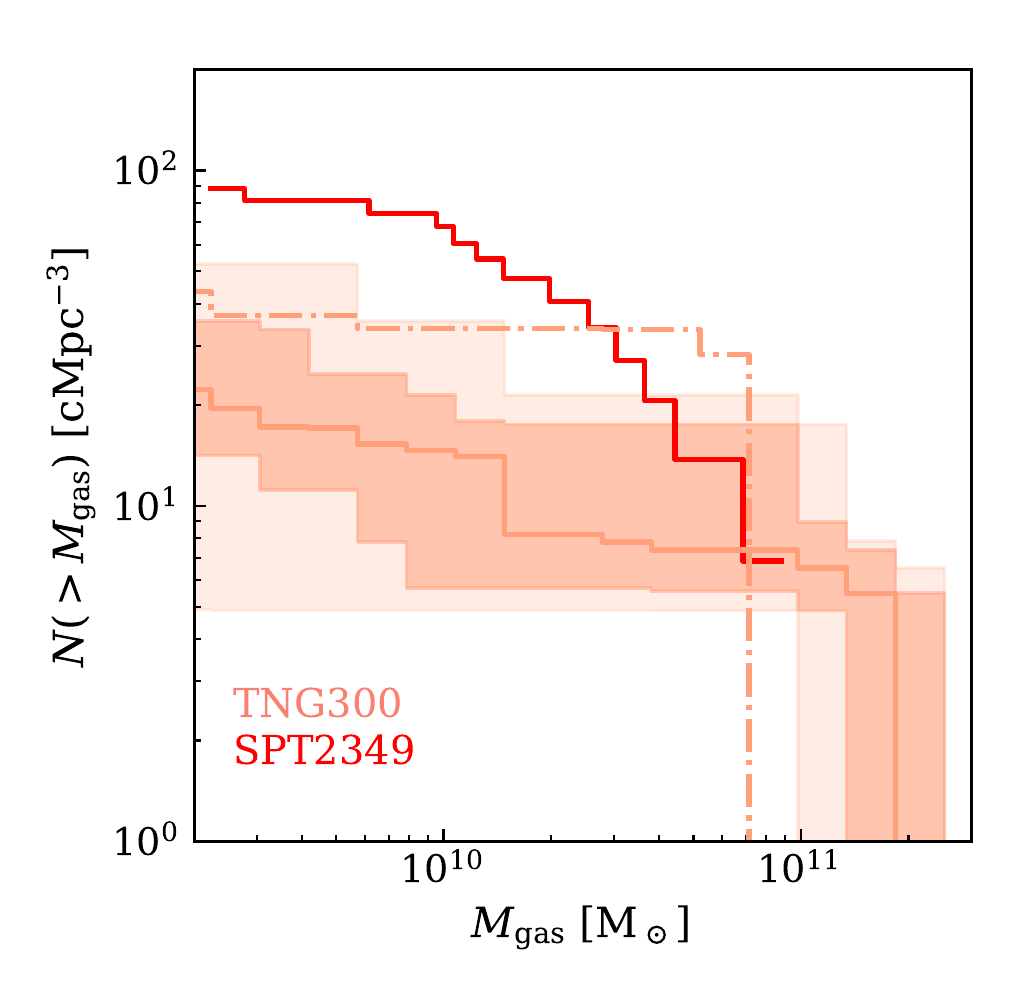}
\caption{Gas mass function of galaxies in the main halo of the 25 most massive protoclusters from TNG300 at $z\,{\simeq}\, 4.3$, with the salmon lines and bands showing the median and 68 and 99.5 per cent ranges, respectively. The dot-dashed line shows the prediction for the TNG300 sample at $z\,{\simeq}\,7$. The red line shows the gas mass function of SPT2349$-$56, obtained using the gas mass and size estimates from H20. } 
\label{fig_gmf}
\end{figure}

Figure~\ref{fig_gmf} shows the gas-mass function of the protocluster galaxies per comoving volume within $R_{500}$, predicted from the TNG300 sample. These are compared with the mass functions obtained for SPT2349$-$56, which is obtained using the gas mass and the size estimates from H20. The abundance of the simulated protocluster galaxies is systematically lower than that coming from the observation, particularly at lower gas mass, although the scatter among individual simulated objects is substantial covering almost a factor of 10 in the number of galaxies per unit volume.

%%%%%%%%%%% figure

%%%%%%% SECTION 7
\section{Predictions at higher redshift}
\label{sec_prediction}

\begin{figure*}
\includegraphics[width=1.0\linewidth]{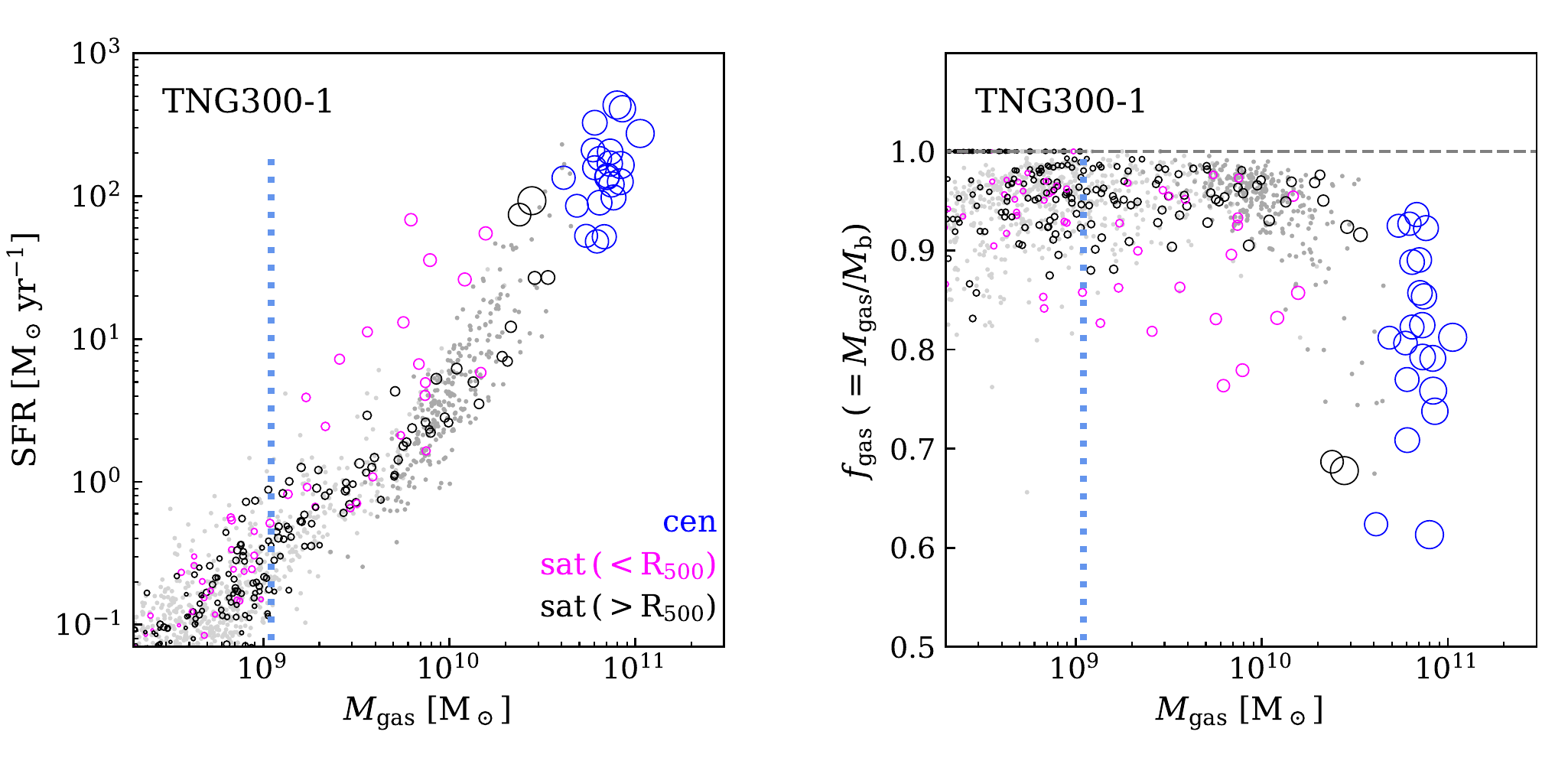}
\caption{SFR (left panel) and gas fraction (right panel) of the galaxies in the main halo of the simulated protoclusters predicted from TNG300 (unfilled circles) at $z\,{\simeq}\, 7$, shown separately for centrals (blue) and for satellites within (magenta) and outside (black) $R_{500}$ from the centre. The dark (light) grey dots show other central (satellite) galaxies from the simulation that do not belong to the sampled protoclusters. The vertical lines indicate the mass limits corresponding to the limit of 100 particles for objects in the simulations (Sect.~\ref{sec_methods}). The sizes of the circles are proportional to the sizes of the subhaloes associated with galaxies, calculated based on the total masses of subhaloes, with the correction for mass loss due to stripping as described in Sect.~\ref{ssec_SFR}. } 
\label{fig_prediction}
\end{figure*}

Simulations allow us to make predictions for the evolution of protoclusters out to even higher redshifts than have yet been seen in the observations. However, with upcoming large and deep millimetre-wave surveys, we should expect that even more distant protoclusters will be discovered.  While several earlier figures in this paper have shown some predictions out to high redshift, Fig.~\ref{fig_prediction} explicitly shows the results for the SFRs and gas mass fractions of protocluster galaxies at a redshift of $z\,{\simeq}\,7$ from TNG300.  It can be seen that the simulations predict that the central galaxies of protoclusters are more distinct from their satellites at higher redshift, and also that field SMGs become less distinct from protocluster galaxies than generally seen at lower redshift. We already mentioned that the SFR profile of protoclusters is predicted to be denser at higher redshift (Fig.~\ref{fig_prof}), broadly consistent with observations. We also find that the integrated SFRs of protoclusters are predicted to be lower by up to an order of magnitude at higher redshift, relative to its peak at $z\,{\simeq}\, 4$--5 (Figs.~\ref{fig_SFH_TNG} and \ref{fig_SFH}). Finally, the expected mass of a progenitor at $z\,{\simeq}\,6$--7 for a Coma cluster-like object is $10^{11}$--$10^{12}\,{\rm M_\odot}$, with a size ($R_{500}$) between 10 and 30\,kpc (Fig.~\ref{fig_MAH}).  Discovery of protoclusters at such redshifts, and the subsequent study of their physical properties, will therefore provide strong tests of whether the simulations contain the necessary star-formation prescriptions.

%%%%%%%%%%% figure

%%%%%%% SECTION 8
\section[discussion]{DISCUSSION}
\label{sec_discussion}

\subsection{Resolution dependence}

As briefly mentioned in Sect.~\ref{sec_model}, the dependence on numerical resolution for some of the model predictions from hydrodynamical simulations remains a fundamental limitation, while the level of this dependence and the rate at which numerical convergence is reached as resolution is increased is highly dependent on {\it both\/} the choice of observables and the scales of interest \citep[e.g.][]{schaye2015, pillepich2018}. The right panel of Fig.~\ref{fig_SFR_Mgas} shows the SFRs of protocluster galaxies at a given gas mass; this is the same as the left panel of the same figure, but using the results from TNG300-3, whose resolution is lower by a factor of $64$ ($4$) in mass (spatial) scale. Note that this resolution difference corresponds to `two levels' of difference, with a factor of $8$ ($2$) difference typically regarded as `one level' of difference in numerical simulations. One can clearly see that the SFRs of galaxies (in particular those with smaller gas mass) predicted from TNG300-3 are significantly lower than from TNG300-1 by more than a factor of 3 ($0.5\,$dex) in some cases. Indeed, the results from the MACSIS simulations shown in the middle panel of the same figure, whose resolution is close to TNG300-3 (with less than `one level' of difference in resolution), are very similar to those from TNG300-3. This may also indicate that the differences in resolution dominate the effects of any difference in the hydro solvers between the two simulations, which can also affect the predictions. Similarly, we find that the predictions from TNG300-2, whose resolution is between TNG300-1 and TNG300-3, are between those from TNG300-1 and TNG300-3 (although the results are not shown). The right panel of Fig.~\ref{fig_fgas} presents the same results for the gas fraction as in the left panel of the figure, but for TNG300-3; this is also shown to be sensitive to the resolution, likely due to the resolution dependence of the SFRs transferred to the stellar masses. All these considerations imply that star formation in high-$z$ protocluster galaxies as predicted by simulations strongly depend on numerical resolution, likely dominating other factors and differences in models that might affect the comparison. This is despite the fact that the details of the sub-grid physics have been tuned to match the properties of typical galaxies (even although tuning to match the properties of average galaxies does not guarantee a match to properties for a particular subset of galaxies, e.g. the SFR-selected protoclusters at high-$z$ studied in this paper), making them important constraints for models, as will be discussed in detail in Sect.~\ref{ssec_limits}. 

One factor that may be responsible for such a strong resolution dependence of SFR from simulations is that the spatial resolution of the simulations is insufficient to resolve tidal torques, which can lead to an underestimation of starbursts driven by mergers or tidal interactions of galaxies \citep{sparre2015, sparre2016}. Indeed, \citet{sparre2016} demonstrated directly that SFR is enhanced by an order of magnitude in zoom simulations of mergers with higher resolutions than parent simulations. As shown in this paper and also in other previous studies \citep[e.g.][]{grand2017, pillepich2018}, the numerical convergence of predictions for star formation has not yet been fully reached. This also naturally leads to a choice to make between so-called weak and strong convergence, as discussed in e.g. \citet{schaye2015}, i.e. whether or not to `re-scale' the parameters of subgrid models when changing resolution. We only briefly discuss this here and refer the reader to the original papers specific to the topic. The idea arises because, as just mentioned, in any simulation relying on subgrid prescriptions, the parameters are just values tuned to match observations at a given resolution. Thus, in principle, there is no logical reason to keep the parameter values the same when changing resolution, even if the models themselves and other components of a simulation remain the same. Weak convergence is considered to be met when the parameters are re-scaled for a given resolution and the predictions with the re-scaled parameters do not change. Strong convergence, on the other hand, is considered to be met when the predictions do not vary with resolution, while the parameters are kept the same. 

Notably, the strong dependence on numerical resolution found in our analysis is in conflict with some findings from \citet{bassini20}. \citet{bassini20} performed zoomed-in simulations of 12 galaxy clusters, selected from a parent dark-matter-only simulation of $1\,h^{-1}{\rm cGpc}$ on a side, all with $M_{200}\,{>}\,8\times10^{14}\,h^{-1}{\rm M_\odot}$ at $z\,{\simeq}\,0$, with a resolution that is similar to TNG300-1 and 10 times better than that of the simulations used by \citet{granato15}. They reported a similar level of discrepancy to our analysis in SFR with observations for both protoclusters and individual galaxies. However, they found that the SFR calculated within a $\simeq 2\,{\rm Mpc}$ aperture is not any higher than that from \citet{granato15}, who used lower-resolution simulations. This may indicate that details of the numerical convergence of the SFR are subject to particular subgrid prescriptions and numerical implementations used in the simulations, as well as redshift (the comparison with \citealt{granato15} in \citealt{bassini20} was made at a lower redshift of $z\,{\simeq}\,2$). Moreover, the resolution is not the only difference between the simulations used in \citet{granato15} and \citet{bassini20}. \citet{bassini20}, instead, used a different prescription for AGN feedback, which might have also altered the prediction with respect to \citet{granato15}. Overall, the origin of the apparently conflicting conclusions remains unclear and needs to be studied further in future. 

So far we have been only discussing the spatial resolution, while, for the specific topic of this paper, the time resolution between simulation snapshots and outputs may also be important. Using SPT2349$-$56 as the initial conditions for an $N$-body simulation, \citet{rennehan2020} found that it takes 100\,Myr or less for its member galaxies to merge and form a single galaxy. Because the timescale for the merger process is only comparable to or less than that between the TNG300 snapshots (and much shorter than that of MACSIS), it is likely that some systems undergoing a rapid collapse with an extremely boosted total SFR are `invisible' in any of the simulation snapshots.

\subsection{Subgrid models for AGN feedback}

Another specific area of uncertainty regarding the model predictions lies in subgrid details implemented for AGN feedback. The absence of AGN feedback would lead, through catastrophic cooling, to prediction of systems that are too massive, with extremely high SFRs to match observation at $z\,{\simeq}\,0$. However, details of how exactly the AGN feeback injects energy and affects the gas are poorly understood, leaving lots of uncertainty in modelling it. For example, most simulations, including the ones used in our analysis, have black holes that are seeded and kept (by `repositioning') at the centres of galaxy systems; this could increase the strength of feedback (i.e. creating stronger outflows), and suppress star formation relative to possibly off-centred or moving black holes in real galaxies or in some other simulations \citep{tremmel2017, tremmel2019}.  However, note that \citet{bassini20} even found that a complete absence of AGN feedback in their simulations does not boost the star-formation efficiency for large gas mass at high redshift to match the SPT2349$-$56 galaxies, while shortening the characteristic timescale of the star formation process does. Details of the repositioning, including the actual timescale over which it happens, requires further study.

\subsection{Uncertainty in abundance}

As we have already mentioned, it is not straightforward to estimate the number density of observed protoclusters because not all of them are detected and selected with the same criteria and also because they are extremely rare objects. For these reasons, the estimated abundance of massive protoclusters quoted in the literature generally ranges over two orders of magnitude, and some studies hint that it may be as low as only one per $10\,{\rm Gpc}^3$ depending on the selection criteria \citep[e.g.][]{wang2020}, which is smaller than the abundance we assumed in our analysis by more than a factor of 10. Although MACIS, the largest simulation in our study, is expected to contain at least a few massive protoclusters, there could still be none, due to chance within Poisson statistics, the specific choice of initial conditions, or a sample selection that does not perfectly match the observational selections.

\subsection{Other limitations}
\label{ssec_limits}

Another caveat of the models is that they are normally designed to yield accurate predictions for average populations, since the observations with which the models are tuned are usually dominated by typical galaxies at any given redshift and mass scale. Additionally, the galaxy relations used to adjust the details of the models (including star-formation recipes and feedback) often concentrate on populations of galaxies selected through something close to stellar mass in the optical.  On the other hand, we have been stressing here the alternative selection (at longer wavelengths) through SFR. Because of this, models may not be tuned in such a way that rare, extreme objects such as protoclusters with high star formation at high-$z$ are represented properly. So perhaps the discrepancies between the simulated and real protoclusters is telling us that the models need to be specifically adapted in the high-SFR regime. Yet another issue with models is degeneracy; most simulations use observations at relatively low redshifts, such as the IMF or scaling relations at $z\,{\simeq}\, 0$, to tune their model parameters, while in practice different evolutionary paths of galaxies may lead to remarkably similar $z\,{=}\,0$ scaling relations. 

Furthermore, comparisons between observations and models are non-trivial, with several issues leading to general limitations. One such issue is selection effects. Protoclusters in this study are normally identified through their extremely high estimated SFRs. Although our analysis tried to minimize selection effects by applying selection cuts in both mass and SFR, as well as by inspecting all individual objects, there could still be a residual selection bias due to the large uncertainty in the number densities of the observed protoclusters (that we used to select the samples). Another, related caveat is a variance in the sampling of populations due to limited volume and realizations, which is purely statistical in nature. Moreover, fluctuation modes with scales larger than the size of the periodic box are lost in simulations. The sampling variance may be a particular issue for the study presented in this paper, where attention is focused on some of the rarest objects in the Universe. 

Finally, observational measurements, and estimates derived from them, are accompanied by uncertainties of their own.  These can be substantial in size, particularly regarding the conversion factors used to compute SFRs and masses from direct observables, including rest-frame FIR fluxes and line-emission strengths. As already mentioned in Sect.~\ref{sec_spt2349}, a different calibration using the dust continuum results in gas-mass estimates higher by about an order of magnitude than gas masses estimated from CO line intensities. Moreover, the $\alpha_{\rm CO}$ conversion factor currently adopted in the literature spans about an order of magnitude \citep[e.g.][]{carilli2013, bolatto2015, aravena2016, bothwell2017}. Conversion factors between different line transitions are another source of uncertainty of the same kind \citep[e.g.][]{bothwell2013, spilker2014}. Additionally, the ratio of SFR to FIR flux density used in studies is usually assumed to be a single value over all galaxies \citep[e.g.][]{barger2014}, while the actual ratio is probably subject to the dust absorption efficiency of each individual galaxy \citep[e.g.][]{safarzadeh2016}, as well as geometrical effects. Lastly, the conversions are often made through factors that are mostly derived based on observations at low redshift (that are largely untested at high redshift), which adds additional uncertainties.

%%%%%%% SECTION 9
\section[summary]{SUMMARY AND CONCLUSIONS}
\label{sec_sum}

In this paper we have presented a comparison between model predictions and observations regarding the properties of star-formation-selected protoclusters, using a high-resolution hydrodynamical cosmological simulation, TNG300, and zoom simulations of $390$ massive clusters, MACSIS, together with two empirical galaxy formation models from \citet{behroozi2013} and \citet{moster2013}. The models chosen have different advantages and caveats of their own, such that they are complementary to each other, allowing us to explore a wide range of limitations that might be included in current models. For comparison with observational data, we have used a compilation of observations of a total of 17 protoclusters from the literature, which are located between redshift 2 and 7, specifically focusing on structures selected through the active star formation of their member galaxies. We paid extra attention to ensure that comparisons are fair and uniform by shifting to quantities in the same apertures (of $5R_{500}$) and by applying a selection cut for simulated galaxies to mimic observations. We found that the difference between using and not using these aperture corrections is significant. We also checked that the conclusions did not change when the selection of (proto)clusters and their main haloes in the simulations was made based on their mass at $z\,{=}\,0$ or their SFR at each redshift. 

Despite uncertainties included in the data arising from different ways of estimating mass, we have found good agreement for the mass-accretion history (MAH) of protoclusters between the models and the data.  By this we mean that most (proto)clusters in the sample lie on the same evolutionary path to become a Coma-like cluster at $z\,{\simeq}\,0$, with a typical total mass of $M_{\rm 200}\,{\simeq}\,10^{13.5-14}{\rm M_\odot}$ at $z\,{=}\,2$ and ${\simeq}\,10^{12.5-13}{\rm M_\odot}$ at $z\,{=}\,4$. This good agreement indicates that the protoclusters from the models that we used in our analysis are representative of the most massive objects seen in the observations. 

So the haloes appear to match what is needed to make high-redshift protoclusters -- but what about their star-formation properties?
To answer this question we investigated the SFHs of simulated protoclusters, and here the simulations are in much poorer agreement with the data. Most strikingly, simulations predict at $z\,{>}\,2$ a total SFR of a few hundred ${\rm M_\odot\,yr^{-1}}$, on average, and up to a few thousand ${\rm M_\odot\,yr^{-1}}$ in protoclusters, while the observational data indicate a higher SFR by a factor of at least 3. From each of the simulations, only the protoclusters with the highest SFRs marginally overlap with the observed protoclusters of the lowest SFRs at $z\,{>}\,2$. Clearly, current simulations do not produce protoclusters that match the observations. Given the good agreement previously found for the MAH, this substantial discrepancy implies a deficiency of star formation in haloes of similar mass. 

Furthermore, the simulated protoclusters are found to have highly centralized star formation, with about 80 per cent (50 per cent) of the total star formation within $5R_{500}$ actually occurring within $R_{500}$ at $z\,{\simeq}\,4$ ($z\,{\simeq}\, 2.5$), which is somewhat higher on average than seen in observations. The simulations agree qualitatively with the observation that the SFR profile becomes even denser with increasing redshift. However, the total SFRs of protocluster as a whole at any given redshift are predicted to be dominated by the star formation occurring outside $5R_{500}$, implying a change in the slope of the SFR profile for a given protocluster. 

The discrepancy in SFR is found not only for entire protoclusters, but also for their individual member galaxies. Simulations lack member galaxies with SFRs greater than ${\rm 10^3\,M_\odot\,yr^{-1}}$, which are confirmed to exist in observations (assuming of course that the data-based SFR estimates are correct). Using SPT2349$-$56 as a case study to further investigate these discrepancies, we find that the instantaneous SFRs of galaxies at a given gas mass are higher by more than an order of magnitude for the SPT2349$-$56 galaxies compared to the simulations. The simulations also predict a much higher gas fraction of about 0.7 for protocluster galaxies, compared to the value of around 0.2 found from the observations. Finally, protocluster galaxies from the simulations show boosted SFRs in denser environments (possibly due to mergers and tidal interactions), whereas those in the observations show no obvious difference relative to SMGs that are not in protoclusters. 

We show that all these discrepancies can, in large part, be traced to the resolution dependence of SFRs in simulations. Simulations with lower resolution, by a factor of $64$ in mass, yield a drop of about a factor of 3 in SFR for galaxies at a given gas mass. One possible explanation is that this may be due to insufficient resolution to resolve tidal torques, which may lead to an underestimation of the star formation triggered by mergers or the tidal interactions of galaxies. Also, the limited time resolution between the simulation snapshots (${>}\,70\,{\rm Myr}$) is lower than the estimated timescale of collapse for a system like SPT2349$-$56 \citep{rennehan2020}, meaning that some overdense systems undergoing rapid coalescence may not be properly captured in any simulation snapshot. Additionally, uncertainty in the estimated abundance of observed protoclusters and sample selection poses an ambiguity in our comparisons. 

Other uncertainties, including the seeding of black holes, assumptions based on low-$z$ observations, and ambiguities in comparing simulated data with direct observables, may also add biases to our comparisons. Also, models are usually tuned to match observations averaged over populations and at $z\,{\simeq}\,0$, making less obvious their predictive power for extreme objects at high redshift, such as the star-formation-selected protoclusters investigated in this paper. Such objects may therefore be important to study in order to refine the sub-grid physics recipes in the simulations, and hence to learn more about the connections between star formation and structure formation in the early Universe. 

Continuing and future millimetre-wavelength surveys are likely to probe even rarer star-forming protoclusters at higher redshifts. We therefore also presented the predictions of simulations for protoclusters out to $z\,{\simeq}\,7$.  Here we find that central galaxies are more distinct relative to satellites, and that galaxies in protoclusters stand out less distinctly compared to star-forming galaxies outside of protoclusters. Moreover, galaxies are predicted to be more gas rich, but with significantly lower star-formation efficiencies, on average, in haloes of a given mass than at lower redshift.  The gathering of larger samples of protoclusters, extending out to higher redshifts, is therefore likely to be even more useful for refining our understanding of star formation within early structures.

\section*{ACKNOWLEDGEMENTS}

We are grateful to Ryley Hill, Scott Chapman and Chris Hayward for many useful discussions and suggestions that significantly contributed to this paper. We acknowledge the referee for his/her constructive comments that improved the paper significantly. We thank Peter Behroozi for providing the data for the empirical model, and George Wang for providing the observational SFRs for the eight SPT protocluster candidates. We acknowledge all groups involved for making the IllustrisTNG simulation data available. The MACSIS simulations used the DiRAC Data Centric system at Durham University, operated by the Institute for Computational Cosmology on behalf of the STFC DiRAC HPC Facility (www.dirac.ac.uk); this equipment was funded by BIS National E-infrastructure capital grant ST/K00042X/1, STFC capital grants ST/H008519/1 and ST/K00087X/1, STFC DiRAC Operations grant ST/K003267/1 and Durham University. DiRAC is part of the National E-Infrastructure. This research was supported by the Natural Sciences and Engineering Research Council of Canada.

\section*{DATA AVAILABILITY}

The data underlying this article will be shared on reasonable request to the corresponding author. The output of the flagship runs of the IllustrisTNG project is publicly available and accessible at www.tng-project.org/data \citep{nelson19b}.

\bibliographystyle{mnras}
\bibliography{protocluster_bib}

\appendix

\section{SFR-based selection of protoclusters and main haloes}
\label{sec_appA}

Selecting the protoclusters and their main haloes from simulations based on their SFR, instead of their mass, may be more appropriate for the comparisons presented here because the protocluster samples from the observations were mainly selected because of their high SFRs. Massive $z\,{=}\,0$ clusters in simulations are shown to have a wide spread in their accretion histories, such as the redshift of their first collapse and the fraction of those with major mergers at high redshift \citep{rennehan2020}. Additionally, the most massive clusters selected at $z\,{=}\,0$ are not necessarily the most massive ones at high $z$. Thus, here we consider two changes in the selection of simulated protoclusters and their main haloes: we choose the 25 protoclusters with the highest total SFRs from the simulation snapshots at each given redshift (instead of selecting the 25 most massive ones at $z\,{=}\,0$, and tracking them across the snapshots), and we also define the `main' haloes to be the ones with the highest SFRs among all haloes of a given protocluster (we continue to use the same definition of a protocluster as in Sect.~\ref{sec_methods}, namely the collection of all progenitors that end up in a common $z\,{=}\,0$ cluster). Note, however, that with this definition we are not tracking the `same' haloes across the snapshots, because the rank in SFR changes between redshifts, thus any trend with redshift cannot be considered as physical evolution. 

\begin{figure*}
\includegraphics[width=0.95\linewidth]{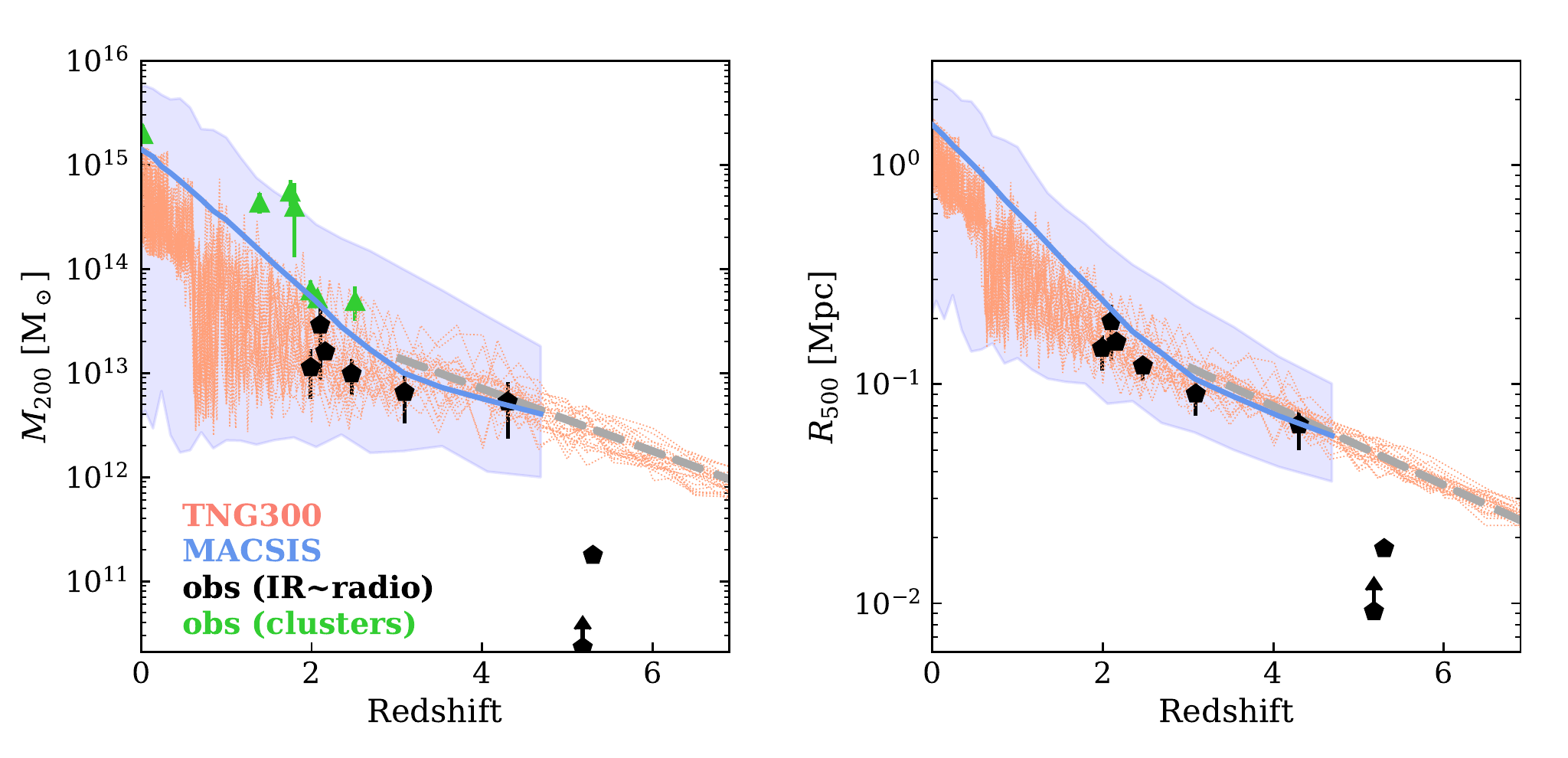}
\caption{Mass (left) and size (right) of the main haloes as a function of redshift, the same as in Fig.~\ref{fig_MAH}, but for the SFR-selected sample (see the text for details of the selection) from TNG300 (salmon lines) and from MACSIS (blue line and band for the median and 99.5 per cent range, respectively). The symbols represent the same observational compilation as in Fig.~\ref{fig_MAH}. We find that the mass and size at $z\,{\gtrsim}\,3$ are described well by a linear function of $\log_{10}M_{200}(z) \,{=}\, 13.15 - 0.30(z-3)$ and $\log_{10}R_{500}(z) \,{=}\, {-}0.92 - 0.18(z-3)$, respectively (grey dashed lines). The scatter around these relations are remarkably small at $z\,{\gtrsim}\,4.5$, meaning that most protoclusters observationally selected for their high SFR would essentially have the same mass and size at a given redshift.}
\label{fig_appA1}
\end{figure*}

Figure~\ref{fig_appA1} shows $M_{200}$ and $R_{500}$ for the main haloes of the protoclusters selected and defined as above, from both the TNG300 and MACSIS simulations. Note that selecting the 25 protoclusters with the highest SFR is not possible for MACSIS, since the zoom simulations contain only objects pre-selected based on their $z\,{=}\,0$ mass. The SFR-based definition of the main halo, on the other hand, is possible and is applied for the MACSIS samples in this figure. The wide spread and substantial fluctuation found in the mass at $z\,{\lesssim}\,2$ is due to the SFH, shown in Fig.~\ref{fig_SFH_TNG}, which makes the SFR-based selection rapidly change the objects it tracks. For both TNG300 and MACSIS, the mass distribution of the main haloes of protoclusters at $z\,{\gtrsim}\,2$ is shown to be much narrower for the SFR-selected samples than for the $z\,{=}\,0$ mass-selected samples. This is mainly because the MAH is vastly different among individual $z\,{=}\,0$ clusters, and thus the 25 most massive clusters picked up at $z\,{=}\,0$ do not perfectly match those picked up at the higher redshifts, even though the mass is still strongly correlated with SFR at high redshift. The MACSIS samples, on the other hand, for which the selection of protoclusters is still based on the $z\,{=}\,0$ mass, show a similar dispersion in mass at high redshift compared to what is seen in Fig.~\ref{fig_MAH}. This indicates that it is the selection of protoclusters, rather than the definition of the main halo, that affects samples and thus the comparisons at high-$z$, which we directly confirm from the simulations. The mass across redshift obtained with the SFR-based selection shows improved agreement with the observations, despite the considerably reduced object-to-object scatter, indicating that the SFR-based selection is a better representation of the observed protoclusters, as expected. Exceptions are HDF\,850.1 and the AzTEC-3 structure, for which the mass estimates are less reliable. The simulations predict that the median mass can be roughly described by two linear relations (in log-space for the mass) with the slope changing at $z\,{\simeq}\,3$. TNG300 predicts that the masses of protoclusters at $z\,{\gtrsim}\,3$ can be described fairly well by a simple linear function as $\log_{10}M_{200}(z) \,{=}\, 13.15 - 0.30(z-3)$. At $z\,{\gtrsim}\,4.5$, the scatter around this relation is very small, about 0.1\,dex, predicting that the masses of observed protoclusters should be more or less the same at a given redshift. Similarly, the size at $z\,{\gtrsim}\,3$ is described well by $\log_{10}R_{500}(z) \,{=}\, {-}0.92 - 0.18(z-3)$, with a scatter of around 0.03\,dex. 

\begin{figure*}
\includegraphics[width=0.95\linewidth]{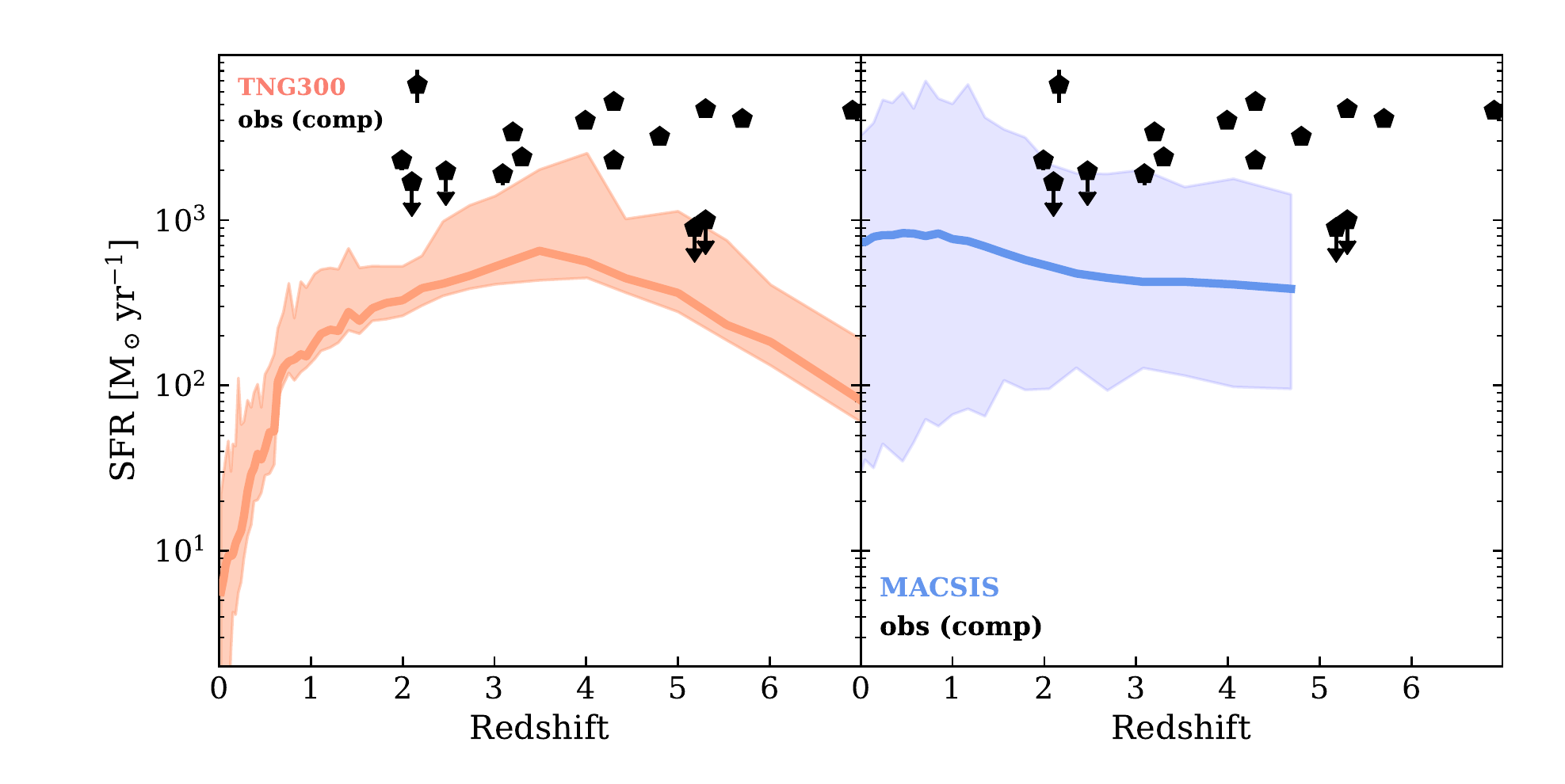}
\caption{Star-formation rate, calculated within $5R_{500}$, of the main haloes selected at a given redshift for the SFR-selected sample from TNG300 (salmon line and band showing the median and 99.5 per cent range, respectively) and from MACSIS (blue line and band showing the median and 99.5 per cent range, respectively). The symbols show the same compilation of observations as in Fig.~\ref{fig_SFH}, as described in Sect.~\ref{sec_obs} and corrected to the aperture of $5R_{500}$ in Sect.~\ref{sec_methods}. }
\label{fig_appA2}
\end{figure*}

Figure~\ref{fig_appA2} shows the same SFH as in Fig.~\ref{fig_SFH}, calculated within $5R_{500}$ of the main halo, but for the SFR-selected samples, compared to the same observations. One can see that the most notable change is the highly elevated lower limit in SFR, and the median trend also changes significantly by up to an order of magnitude. The upper limit, however, barely shifts, thus retaining a significant discrepancy with the observations, indicating that the protoclusters with the highest SFRs at high redshift are among the most massive ones at $z\,{=}\,0$. Finally, we directly check that the discrepancy with the observations remains at a similar level even when selecting the highest SFR haloes (regardless of their association with clusters, even allowing for the selection of multiple haloes within a single protocluster as long as they have the highest SFR at a given redshift) in each snapshot from TNG300. This simply shows that there is no object in the simulations with an SFR high enough to match those of the observed protoclusters. 

%%%%%%%%%%% figure

%%%%%%%%%%% figure

\label{lastpage}

\end{document}